\documentclass[12pt]{article}
\usepackage[a4paper,top=2.5 cm,bottom=2.5cm,left=2.5cm,right=2.5cm]{geometry}

\usepackage[utf8]{inputenc}				
\usepackage{mdframed}					
\usepackage{graphicx}					

\usepackage{caption}
\usepackage{subcaption}

\graphicspath{{figures/}}
\usepackage{xspace}
\usepackage{amsmath}

\usepackage{amsthm}
\theoremstyle{remark}

\usepackage{mathtools}
\usepackage{enumitem}
\usepackage[super]{nth}

\usepackage[backend=bibtex, sorting=none, style=numeric, citestyle=numeric]{biblatex}

\DeclareNameAlias{sortname}{last-first}

\usepackage{hyperref}

\mathtoolsset{showonlyrefs}

\usepackage{comment}

\usepackage{float}

\usepackage{enumitem}

\usepackage{color}

\usepackage{physics}
\newcommand{\underbrset}[2]{\underset{#1}{\underbrace{#2}}}

\usepackage{amssymb}
\usepackage{upgreek}

\newcommand{\bbR}{\ensuremath\mathbb{R}} 

\newcommand{\vect}[1]{%
	{\boldsymbol{\mathbf{%
				\mathit{#1}
	}}}
}

\newcommand{\ctens}[1]{%
	{\boldsymbol{\mathbf{%
				\mathit{#1}
	}}}
}

\newcommand{\SDvect}[1]{%
{\mathbf{\mathbf{%
            \mathbf{#1}
    }}}
}

\newcommand{\SDtens}[1]{%
{{
{\mathbf{%
            \mathbf{#1}
    }}}}
}

\newcommand{\ctensd}[1]{\mathsf{{#1}}}

\makeatletter
\newcommand{\printfnsymbol}[1]{%
	\textsuperscript{\@fnsymbol{#1}}%
}
\makeatother

\usepackage{authblk}
	
\usepackage{color}
	
	\title{A time multiscale based data-driven approach in cyclic elasto-plasticity}
	
	\author[,1]{Sebastian Rodriguez\thanks{Corresponding author}\thanks{These authors equally contributed}}
	\author[,1,2]{Angelo Pasquale\printfnsymbol{1}\printfnsymbol{2}}
	\author[3]{Khanh Nguyen}
	\author[2,4]{Amine Ammar}
	\author[1,4]{Francisco Chinesta}
	
	\affil[1]{ESI Group Chair @ PIMM Lab, ENSAM Institute of Technology, 151 Boulevard de l'Hôpital, F-75013, Paris, France, \href{mailto:angelo.pasquale@ensam.eu}{angelo.pasquale@ensam.eu},
		\href{mailto:sebastian.rodriguez_iturra@ensam.eu}{sebastian.rodriguez\_iturra@ensam.eu},
		\href{mailto:francisco.chinesta@ensam.eu}{francisco.chinesta@ensam.eu}\vspace{0.5cm}}
	\affil[2]{ESI Group Chair @ LAMPA Lab, ENSAM Institute of Technology, 2 Boulevard du Ronceray BP 93525, 49035 Angers cedex 01, France, \href{mailto:angelo.pasquale@ensam.eu}{angelo.pasquale@ensam.eu},
		\href{mailto:amine.ammar@ensam.eu}{amine.ammar@ensam.eu}\vspace{0.5cm}}
	\affil[3]{Escuela Técnica Superior de Ingeniería Aeronáutica y del Espacio, Universidad Politécnica de Madrid, Pza. Cardenal Cisneros, 28040, Madrid, Spain, \href{mailto:khanhnguyen.gia@upm.es}{khanhnguyen.gia@upm.es}\vspace{0.5cm}}
	\affil[4]{CNRS@CREATE LTD, 1 Create Way, \#08-01 CREATE Tower, Singapore 138602, Singapore}
	
	\date{}
	
	\makeatletter
	\newcommand*{\toccontents}{\@starttoc{toc}}
	\makeatother

\usepackage{bbm}
\usepackage{tikz,pgfplots}

\usetikzlibrary{decorations.fractals}

\pgfplotsset{compat=newest}
\usetikzlibrary{plotmarks}
\usetikzlibrary{arrows}
\usepgfplotslibrary{patchplots}
\usepackage{grffile}
\pgfplotsset{plot coordinates/math parser=true}
\usepackage{standalone}

\usetikzlibrary{arrows.meta,graphs,decorations.pathmorphing,decorations.markings} 

\usepackage{titlesec}
\usepackage[titletoc,toc,title]{appendix}

\DeclareMathSymbol{\mlq}{\mathord}{operators}{'134}
\DeclareMathSymbol{\mrq}{\mathord}{operators}{'42}

\usepackage{parskip}

\usepackage{algpseudocode,algorithm,algorithmicx}

\algrenewcommand\algorithmicrequire{\textbf{Require:}}
\algrenewcommand\algorithmicensure{\textbf{Postcondition:}}
\algnewcommand\algto{\textbf{ to }}

\usepackage{bbold} 

\begin{document}						

		
		\maketitle
		
	\begin{abstract}
            Within the framework of computational plasticity, recent advances show that the quasi-static response of an elasto-plastic structure under cyclic loadings may exhibit a time multiscale behaviour. In particular, the system response can be computed in terms of time microscale and macroscale modes using a weakly intrusive multi-time Proper Generalized Decomposition (MT-PGD). In this work, such micro-macro characterization of the time response is exploited to build a data-driven model of the elasto-plastic constitutive relation. This can be viewed as a predictor-corrector scheme where the prediction is driven by the macrotime evolution and the correction is performed via a sparse sampling in space. Once the nonlinear term is forecasted, the multi-time PGD algorithm allows the fast computation of the total strain. The algorithm shows considerable gains in terms of computational time, opening new perspectives in the numerical simulation of history-dependent problems defined in very large time intervals.\\

	\end{abstract}
	
	
	\section{Introduction}
    Due to the dramatically long duration of the phenomenon and the stringent requirements on the grid granularity, direct simulations of structures subject to a high number of loading cycles remains a real challenge. For instance, standard numerical techniques fail in simulating the fatigue life, representing a major design issue in various fields of applications such as aircraft, auto parts, railways and jet engines, among many others \cite{SO+20, Zim18}. 
    
    One of the reasons of the excessive complexity stands in the history-dependent behaviours which require the reconstruction of the whole past history \cite{hist-dep-0,hist-dep-1,hist-dep-2,hist-dep-3,hist-dep-4,hist-dep-5}. Indeed, when this is combined with fine spatial meshes and very long time horizons, the computational complexity leads to cost-prohibitive simulations and to the necessity of adopting suitable simplified models. In this sense, the main motivation beyond this work is to perform a further step towards the direct simulation of such problems.
    
    In the framework of cyclic elasto-plasticity, an approach based on the Proper Generalized Decomposition (PGD) is recently been proposed, obtaining a time multiscale representation of the system response \cite{pgd-multiscale-3}. 
    
    In this case, by denoting with $\mathcal{L} (\bullet)$ a generic nonlinear differential operator involving the space derivatives, the addressed quasi-static problem can be written as
    \begin{equation}
    	\label{eq:generic-problem}
    	\mathcal{L} (u (\vect{x}, t)) = f(\vect{x}, t),
    \end{equation}
    where the time dependence is associated to the cyclic loading $f(\vect{x}, t)$. The nonlinear operator $\mathcal{L}$ is decomposed additively into a linear and a nonlinear part, as $\mathcal{L} = \mathcal{L}_{\text{l}} + \mathcal{L}_{\text{nl}}$. If the superscript $(l)$ tracks the nonlinear iteration, problem \eqref{eq:generic-problem} can be linearized as
    \begin{equation}
    	\label{eq:generic-problem-linearized}
    	\mathcal{L}_{\text{l}} (u^{(l)} (\vect{x}, t)) = f(\vect{x}, t) - \mathcal{L}_{\text{nl}} (u^{(l-1)} (\vect{x}, t)),
    \end{equation}
    whose solution may be computed in the multi-time form \cite{pgd-multiscale-3,pgd-multiscale-2} 
    \begin{equation}
    	\label{eq:space-multimodes}
    	u^{(l)}(\vect{x}, t) \approx u^{(l)}(\vect{x}, \tau, T) = \sum_{k} U^\vect{x}_k(\vect{x}) \sum_j U^\tau_{k, j}(\tau) U^T_{k, j}(T).
    \end{equation}
	where $\tau$ denotes the time microscale variable and $T$ the macroscale one.
    
    This task is computationally cheap using the standard PGD
    constructor \cite{PGD,PGD-1,PGD-2,pgd-space-time,pgd-space-time-1}, even when considering parameters \cite{PGD-param,PGD-param-1,PGD-param-2,PGD-param-3}, and becomes even faster when making use of multi-time separated representations in equation \eqref{eq:generic-problem-linearized}, that is
    \begin{equation}
    	\label{eq:space_multi-time_rhs}
    	f(\vect{x}, t) \approx \sum_{j} F^\vect{x}_j(\vect{x}) F^\tau_j(\tau) F^T_j(T),\quad \mathcal{L}_{\text{nl}}(u^{(l-1)}(\vect{x}, t)) \approx \sum_{j} L^\vect{x}_j(\vect{x}) L^\tau_j(\tau) L^T_j(T).
    \end{equation}
    Such expressions may be obtained, among other possibilities, via the higher-order SVD (HOSVD) \cite{HOSVD,fastHOSVD} or the PGD \cite{PGD}.
    
    However, as pointed out in \cite{pgd-multiscale-3}, the calculation of $\mathcal{L}_{\text{nl}}(u^{(l-1)}(\vect{x}, t))$ becomes a tricky issue when
    \begin{equation}
    	\label{eq:nonlinear-term}
    	\mathcal{L}_{\text{nl}}(u^{(l-1)}(\vect{x}, t)) = \mathcal{N}( u^{(l-1)}(\vect{x}, s); s \leq t )
    \end{equation}
	where $\mathcal{N}$ denotes a nonlinear operator. According to \eqref{eq:nonlinear-term}, the nonlinearity is local in space but history-dependent in time, as encountered in elasto-plastic behaviors in solid mechanics. For instance, keeping the same notation of \cite{pgd-multiscale-3} where hardening plasticity is considered, a nonlinear operator $\mathcal{N}$ acts on the total strain tensor and on the effective plastic strain up to the final time $T_f$, that is
    \begin{equation}
    	\label{eq:non-linear-term-hist}
    	\ctens{\varepsilon}^{p, (l - 1)} =  \mathcal{N}(\ctens{\varepsilon}^{(l - 1)}, \bar{\varepsilon}^p_{T_f}).
    \end{equation} 
    Specifically, the evaluation of \eqref{eq:non-linear-term-hist} requires the reconstruction over the whole past history since
    \begin{equation}
    	\bar{\varepsilon}^p_{t} = \int_0^{t} \sqrt{\frac{2}{3} \dot{\ctens{\varepsilon}}^p : \dot{\ctens{\varepsilon}}^p } \dd{s}.
    \end{equation}
    Thus, the construction of the right-hand side entails two main difficulties common to all standard discretization techniques: (i) because of the behavior locality, the nonlinear term must be evaluated at each location $\vect{x}$ used for discretizing equation \eqref{eq:generic-problem-linearized}; (ii) the nonlinear term must be evaluated along the whole (long) time interval with the resolution enforced by the fastest physics, that is $\tau$. These requirements of course compromise the solution of problems defined over large time intervals $I = (0, T_f)$.
    
    Different works have been proposed to overcome such limitations, most of them applied to the LArge Time INcrement (LATIN) method \cite{ladeveze1985famille} together with the PGD. For instance, several multiscale approximations have been developed in \cite{cognard1993large,Arzt1994,bhattacharyya2018model,bhattacharyya2018latin,bhattacharyya2018multi,bhattacharyya2019kinetic}, where the computational time is reduced via an interpolation of the solution at different time scales. Other works rely on some hyper-reduction techniques, such as the Reference Point Method \cite{capaldo2015new,capaldo2017reference} or the extension of the gappy-POD technique to the space-time domain \cite{pgd-multiscale-2}, but these were restricted for nonlinear behaviors at internal variables, which is not the case of history-dependent behaviors.
     
    In the present work, the authors propose to overcome such limitations by combining the PGD-based time multiscale representation proposed in \cite{pgd-multiscale-3} with suitable data-driven techniques. Indeed, the analyzed problem presents two time scales, being one related to the characteristic time of the structure and the other to the system response at the level of the cycle. Moreover, the procedure can be generalized to more than two scales, if needed.  
    
    The multiscale approach here proposed does not require time scale separation, contrarily to time homogenization techniques. The term separation in this work is, actually, meant in the context of separation of variables (i.e., PGD-like). The two time scales are coexisting within the formulation meaning that kinematics and mechanics variables are computed simultaneously along the micro and macro scales. 
    
    To deal with long-term simulations involving a high number of cycles and history-dependent nonlinear constitutive relations, a predictor-corrector scheme is here proposed. This is usual also in cycle-jumping techniques, such as \cite{cycle-jump-0, cycle-jump-1, cycle-jump-2, cycle-jump-3}. In fact, in usual cycle-jumping methods, the extrapolated state is employed as the initial state for future finite element simulations, which are used as reference within the correction step. The drawback of such approach is that, being incremental in the predictions, the committed error is accumulated. 
    
    On the contrary, the predictor-corrector scheme here proposed is not incremental. This is achieved (a) treating exclusively the macrotime functions via the predictor-corrector scheme; (b) accounting for the spatial functions through sparse sampling and data completion techniques; (c) assuming unchanged the microtime functions and, afterwards, correcting them via a successive enrichment.
    
    Denoting with $N_{\vect{x}}$ and $N_t$ the number of space and time degrees of freedom, standard techniques have a complexity scaling as $\mathcal{O}(N_{\vect{x}}N_t)$, where $\mathcal{O}$ is the usual symbol for the asymptotic complexity. The overall computational complexity of the proposed procedure scales as $\mathcal{O}(N_{\vect{x}} + N_\tau + N_T)$, where $N_t = N_\tau N_T$, implying interesting gains.
    		
    As a last introductory comment, the strategy ensures the equilibrium globally in space and time. All the stages of the procedure are based on iterative schemes whose solutions' quality is determined and, if necessary, enhanced according to suitable convergence criteria, guaranteeing robustness.
    
    The paper outline is the following. Section \ref{sec:general} presents a general description of the proposed algorithm. Section \ref{sec:dd-model} enters in the details of all the methods exploited to build the data-driven model. Section \ref{sec:application-case} shows the results and computational gains over a specific test case. Finally, section \ref{sec:conclusion} provides conclusions and perspectives.
    
    \section{Theoretical and numerical framework}
    \label{sec:general}
    
    \subsection{Problem statement}
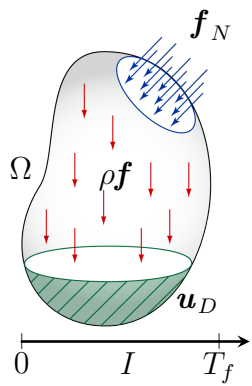
\begin{figure}[!ht]
    	\centering
    	\begin{tikzpicture}[line cap=round,line join=round,>=stealth,scale=1.0]
    		\node[anchor=south west, inner sep=0pt, outer sep=0pt] at (0,0) {\includegraphics[width=2.5cm]{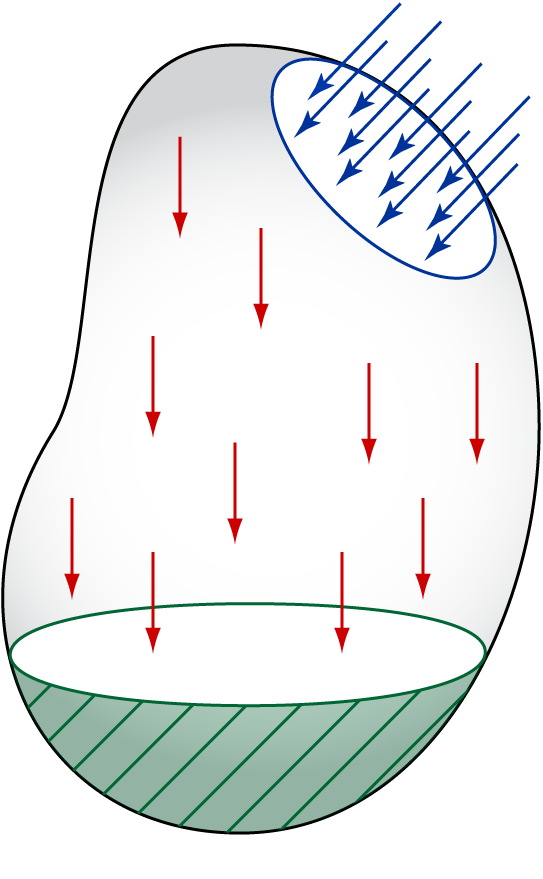}} ;
    		\draw (0.28,2.3) node [left] {\color{black}$\Omega$} ;
    		\draw (2.5,4.2) node [] {$\vect{f}_{N}$} ;
    		\draw (2.3,0.5) node [] {$\vect{u}_{D}$} ;
    		\draw (1.25,2.2) node [] {$\rho \vect{f}$} ;
    		\draw [->][black][line width=1pt][cap=round] (0,0) -- (3.0,0);
    		\draw (0,0) node[] {\color{black}$\scriptscriptstyle{|}$} ;
    		\draw (0,0) node[below] {\color{black}$0$} ;
    		\draw (2.6,0) node[] {\color{black}$\scriptscriptstyle{|}$} ;
    		\draw (2.6,0) node[below] {\color{black}$T_f$} ;
    		\draw (1.4,0) node[below] {\color{black}$I$} ;
    	\end{tikzpicture}
    	\caption{Mechanical problem under study.}
    	\label{fig:ref_prob}
    \end{figure}

    The reference problem consists of an elasto-plastic structure occupying the spatial region $\Omega$ and subject to a cyclic loading $\vect{f} = \vect{f}(\vect{x}, t)$ applied over the time interval $I = (0, T_f)$, as sketched in figure \ref{fig:ref_prob}.
    
    The unknowns are the displacement field $\vect{u} (\vect{x}, t)$ and the stress field $\ctens{\sigma} (\vect{x}, t)$, with $(\vect{x}, t) \in \Omega \times I$, satisfying
    \begin{equation}
    	\label{eq:model-problem}
    	\begin{dcases}
    		\div{\ctens{\sigma}} = \vect{f} & \text{in } \Omega \times I \\
    		\vect{u} = \vect{u}_D & \text{on } \partial \Omega_D \times I \\
    		\ctens{\sigma} \cdot \vect{n} = \vect{f}_N & \text{on } \partial \Omega_N \times I \\
    		\vect{u} = \vect{u}_0 & \text{in } \Omega \times \{0\}.
    	\end{dcases}  
    \end{equation}
	Using standard notations, $\vect{u}_0$ is the initial condition, $\vect{u}_D$ is a prescribed displacement on $\partial \Omega_D$ and $\vect{f}_N$ is a prescribed traction (per unit deformed area) on $\partial \Omega_N$.
    
    Moreover, $\ctens{u}$ and $\ctens{\sigma}$ verify the elasto-plastic constitutive relation
    \begin{equation}
    	\label{eq:Hooke}
    	\ctens{\sigma} = \ctensd{C} : (\ctens{\varepsilon} - \ctens{\varepsilon}^p)
    \end{equation}
    with $\ctensd{C}$ the fourth-order stiffness tensor, $\ctens{\varepsilon} = \nabla^s {{\vect{u}}}$ the total strain tensor ($\nabla^s(\bullet)$ being the symmetric gradient operator), $\ctens{\varepsilon}^p$ the plastic strain tensor and : referring to the tensor product twice contracted.
    
    \subsection{Numerical framework}
    
\begin{figure}[H]
    	\centering
    	\includegraphics[width=0.5\textwidth]{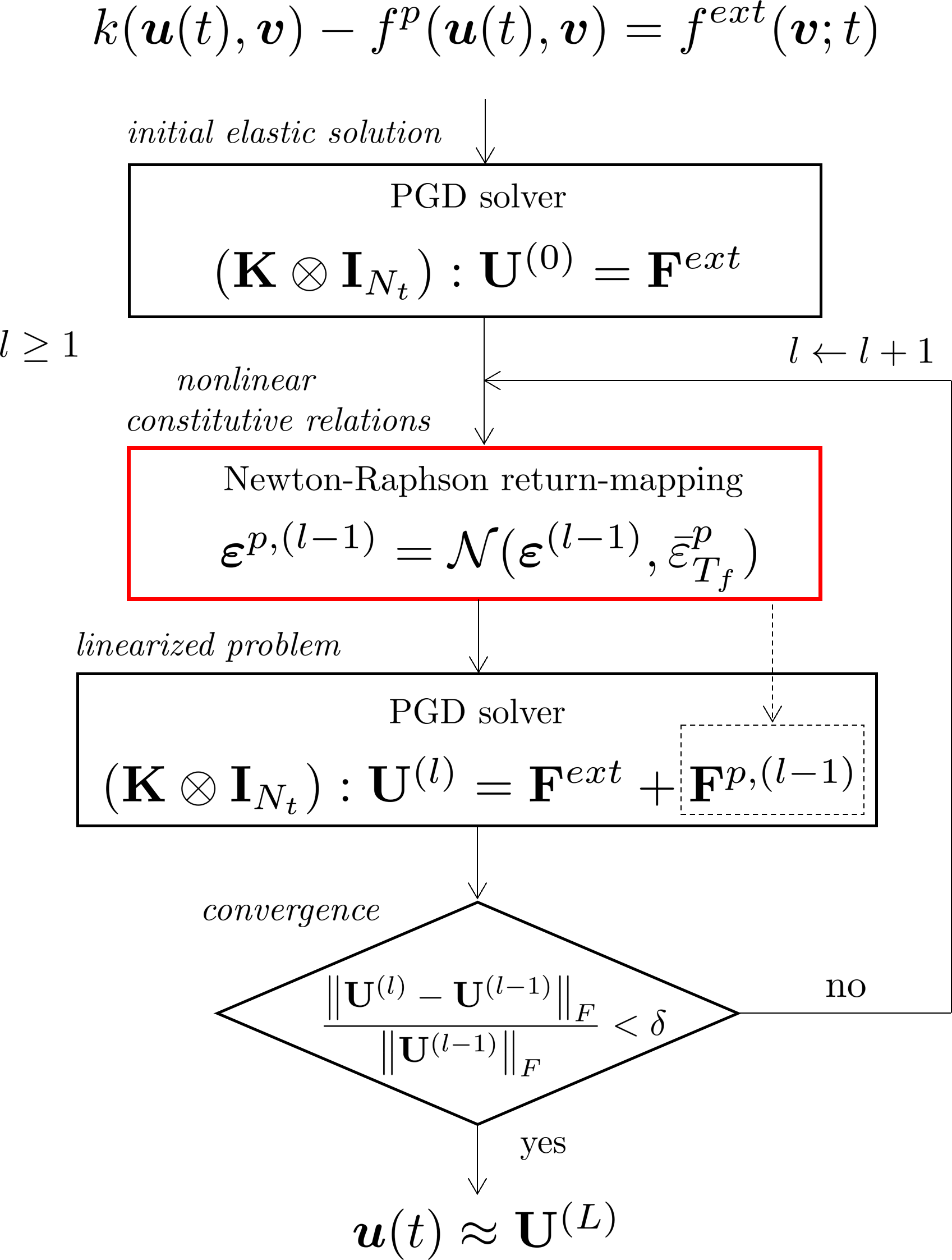}
    	\caption{PGD solving scheme for elasto-plasticity.}
    	\label{fig:simulation-up-to-K}
    \end{figure}

    In terms of numerical algorithms, this work considers the PGD-based procedure proposed in \cite{pgd-multiscale-3}, whose solving scheme is recalled in figure \ref{fig:simulation-up-to-K}.
    
	As detailed in \cite{pgd-multiscale-3}, the first equation of the flowchart in figure \ref{fig:simulation-up-to-K} is the weak form of problem \eqref{eq:model-problem}, based on defining the bilinear and linear forms
	\begin{equation}
		\begin{dcases}
			k(\vect{u}, \vect{v}) = \int_{\Omega} \ctens{\varepsilon} ({\vect{v}}) :  \ctensd{C} : \ctens{\varepsilon} (\vect{u}) \dd{{\vect{x}}} \\
			f^{ext}(\vect{v}; t) = \int_{\Omega} \vect{f}(t) \cdot \vect{v} \dd{{\vect{x}}} + \int_{\partial \Omega_N} \vect{f}_N (t) \cdot {\vect{v}} \dd{{\vect{\gamma}}},
		\end{dcases}
	\end{equation}
	as well as the nonlinear term accounting for the plastic strain
	\begin{equation}
		f^{p}(\vect{u}, \vect{v}) = \int_{\Omega} \ctens{\varepsilon} ({\vect{v}}) : \ctensd{C} : \ctens{\varepsilon}^p (\vect{u}) \dd{{\vect{x}}}.
	\end{equation}
	Once the elastic solution is computed using the PGD solver, an iterative process starts, requiring two steps at each iteration $l \geq 1$:
	\begin{enumerate}
		\item (\textit{state-updating}) The evaluation of the (nonlinear) history-dependent constitutive relations via the elastic predictor/return-mapping procedure
		\begin{equation}
			\label{eq:state-updating}
			\ctens{\varepsilon}^{p, (l - 1)} =  \mathcal{N}(\ctens{\varepsilon}^{(l - 1)}, \bar{\varepsilon}^p_{T_f}),
		\end{equation} 
		where $\mathcal{N}$ represents the nonlinear operator depending on the total strain tensor and on the effective plastic strain up to the final time $T_f$, $\bar{\varepsilon}^p_{T_f}$.  	
		
		\item (\textit{linearized problem}) Re-imposition of the equilibrium
		\begin{equation}
			\label{eq:weak-form-linearized}
			k(\vect{u}^{(l)}(t), \vect{v}) = f^{ext}(\vect{v}; t) + f^{p}(\vect{u}^{(l-1)}(t), \vect{v}),
		\end{equation}
		by means of the PGD solver.
	\end{enumerate}
   The evaluation of the nonlinear constitutive relations (corresponding to the red box in figure \ref{fig:simulation-up-to-K} becomes unfeasible, due to memory and computational issues, when addressing a high-number of cycles. The novel contribution of this work stands in alleviating the computational cost of this step through a data-driven modeling of the nonlinear relations. Moreover, its originality comes from the usage of a time multiscale characterization to build efficiently such model, as detailed in the next section.

    \section{Multiscale-based data-driven modeling}
    \label{sec:dd-model}
	
	Let us suppose that the nonlinear evaluation \eqref{eq:state-updating} can be performed up to $K \ll N$ cycles. This allows to compute the nonlinear terms $ \boldsymbol{\varepsilon}^{p, (l-1)}$ over the space-time domain $\Omega \times I_K$, with $I_K = (0, T_K]$, where $T_K$ denotes the endpoint of the $K$-th loading cycle. Denoting with $T_N$ the endpoint of the $N$-th loading cycle, the aim of the data-driven modeling is to forecast the nonlinear term over $\hat I = (T_K, T_N]$ without additional computational costs\footnote{Similarly, the hat $\hat{\bullet}$ notation will be reserved for the predicted quantities over $\hat I$.}.
	
\begin{figure}[ht]
		\centering
		\includegraphics[width=0.65\textwidth]{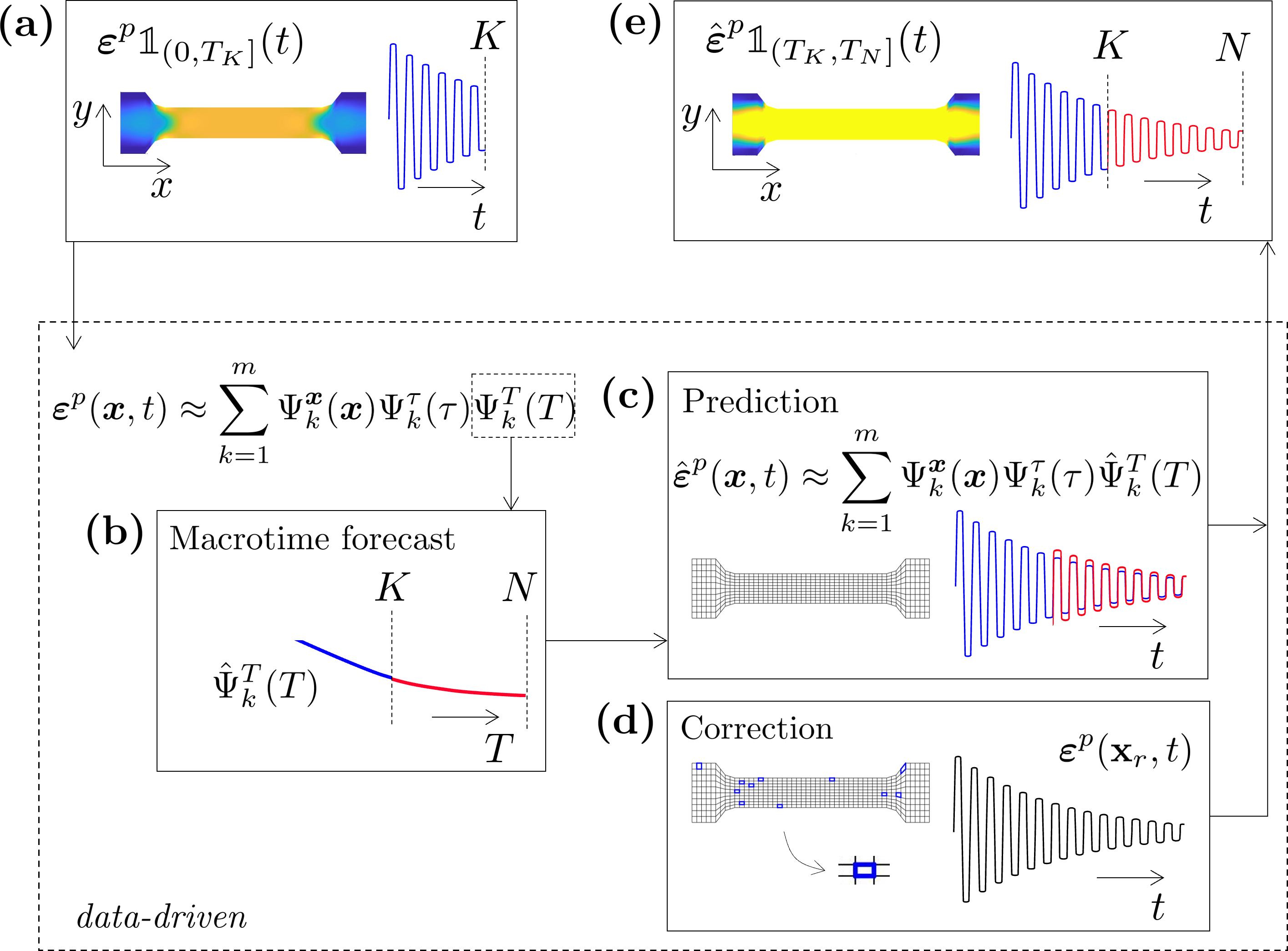}
		\caption{Workflow for the nonlinear constitutive relations prediction.}
		\label{fig:overal-procedure}
	\end{figure}

	Exploiting the multi-time PGD constructor \cite{pgd-multiscale-3,pgd-multiscale-2,pgd-multiscale}, the suggested strategy starts by decomposing the space-time evolution of $\boldsymbol{\varepsilon}^{p, (l-1)}$ in slow and fast time dynamics, via the separated approximation
	\begin{equation}
		\label{eq:eps-space-multimodes}
		\boldsymbol{\varepsilon}^{p, (l-1)} \approx \sum_{k = 1}^m \Psi^{\vect{x}}_k(\vect{x}) \Psi^{\tau}_k(\tau) \Psi^{T}_k(T).
	\end{equation}
	In approximation \eqref{eq:eps-space-multimodes}, a generic function of the microscale $\Psi_k^{\tau}(\tau)$ exhibits a complex highly nonlinear behaviour due to the plasticity occuring over the short scale. On the contrary, a function $\Psi_k^{T}(T)$ of the macroscale is characterized by a really smooth evolution, enabling the accurate and efficient prediction of the long-term evolution. The macrotime predictions are then inserted into a predictor-corrector workflow, as illustrated by the scheme\footnote{In the scheme the superscript $(l-1)$ has been dropped for notational simplicity.} in figure \ref{fig:overal-procedure}.
	
	The workflow consists of five main blocks: (a) performing the nonlinear evaluations up to $T_K$ and computing its multi-time approximation; (b) forecasting the macrotime evolution; (c) predicting the nonlinear response up to $T_N$ using the macrotime forecast; (d) correcting the prediction integrating the nonlinear relations in a few spatial locations; (e) considering the predicted-corrected nonlinear evolution to assemble the linearized problem up to $T_N$.
	
	The following sections explain in detail the predictor-corrector scheme, corresponding to the steps (c) and (d).

    \subsection{Predictor}
    \label{subsec:HODMD}
    
    The predictor is built separately for each macrotime mode $v = \Psi_k^T(T)$, $k = 1, \dots, m$ (scalar-valued function), whose corresponding snapshot (time series) can be written as $\SDvect{v} = (v_1, \dots, v_{N_T}) \in \bbR^{N_T}$. The number of data points coincides with the number of macrodofs $N_T$ and the sampling interval is the macro time step $\Delta T$.
    
    Exploiting only the macro functions has several computational advantages. A few of them are listed here below.
   	\begin{enumerate}
   		\item The size of the analyzed snapshots is reduced. If $N_\tau$ is the number of dofs along the microscale and $N_T$ the number of dofs along the macro one, the length of the time signals reduces from $N_t = N_TN_\tau$ encountered in classical time marching schemes to $N_T$.
   		\item The smooth behavior along the macroscale entails further compression of the snapshots, guaranteeing more memory savings. Indeed,
   		\begin{enumerate}
   			\item the macro modes may be well characterized by means of a few shape parameters $p$ allowing highly-accurate reconstructions (e.g., low-order polynomials) of the signal over all the steps $N_T$;
   			\item a resampling of the macro modes based on $N_T' \ll N_T$ steps will not loose accuracy in the approximation, since all the high frequencies are tracked by the micro modes.
   		\end{enumerate} 
   		\item Forecasting along the macroscale is a much easier task for any time integrator, since all the patterns and highly nonlinear evolution are delegated to the microscale modes. 
   	\end{enumerate}
	
	Among many other possibilities \cite{DMD-time-series,DMD-time-series-1,DMD-time-series-2,DMD-time-series-3,time-integrator}, this work adopts the higher-order DMD for the time series forecasting. The dynamic mode decomposition (DMD) \cite{DMD} is a well known snapshots-based technique allowing to extract relevant patterns in nonlinear dynamics, closely related to the Koopman theory \cite{koopman, koopman-1, koopman-2}. The higher-order DMD (HODMD) is an extension of the former, which considers time-lagged snapshots \cite{HODMD,HODMD-1}. This technique is particularly attractive for the purposes of this work due to its ability of allowing rich extrapolations involving nonzero decaying rates \cite{HODMD}.
    
    The algorithm beyond the HODMD is also called DMD-$d$ algorithm, since it considers $d$-lagged elements. For $d \geq 1$ fixed hyper-parameter, this means that the following higher-order Koopman assumption is made \cite{HODMD}
    \begin{equation}
    	\label{eq:HOkoopman}
    	v_{j + d} \approx c_{1} v_{j} + c_{2} v_{j + 1} + \cdots + c_d v_{j + d - 1},
    \end{equation}
	which is rewritten in terms of standard Koopman assumption as
	\begin{equation}
		\label{eq:koopman-1}
		\tilde{\SDvect{v}}_{j + 1} \approx \tilde{\SDvect{R}} \tilde{\SDvect{v}}_j,
	\end{equation}
	involving enlarged snapshots and (unknown) Koopman matrix
	\begin{equation}
		\label{eq:koopman-snapshots-1}
		\tilde{\SDvect{v}}_j = \begin{pmatrix}
			v_j \\
			v_{j + 1} \\
			\vdots \\
			v_{j + d - 2} \\
			v_{j + d - 1}
		\end{pmatrix} \in \bbR^d,\quad \tilde{\SDvect{R}} = \begin{pmatrix}
			0 & 1 & 0 & \cdots & 0 & 0 \\
			0 & 0 & 1 & \ddots & \vdots & \vdots \\
			\vdots & \vdots & \ddots & \ddots & 0 & 0 \\
			0 & 0 & \cdots & 0 & 1 & 0  \\
			c_1 & c_2 & c_3 & \cdots & c_{d - 1} & c_d 
		\end{pmatrix} \in \bbR^{d \times d},
	\end{equation}
	with $1 \leq j \leq N_T - d$.
	
	Once the HODMD-based models for the macroscale modes $\{\Psi_k^T(T)\}_{k=1}^m$ are trained, they give the predictions $\{\hat \Psi_k^T(T)\}_{k=1}^m$ over $\hat I$. Re-using the microscale and spatial modes from \eqref{eq:eps-space-multimodes}, the nonlinear response is predicted over $\hat I$ as
	\begin{equation}
		\label{eq:3PGD-multiscale-predicted}
		\hat{\boldsymbol{\varepsilon}}^{p, l - 1}(\vect{x}, t)\mathbb{1}_{\hat I}(t) = \sum_{k = 1}^m \Psi^{\vect{x}}_k(\vect{x}) \Psi^{\tau}_k(\tau) \hat{\Psi}^T_k(T).
	\end{equation}
	
\begin{figure}[H]
		\centering
		\includegraphics[width=0.5\textwidth]{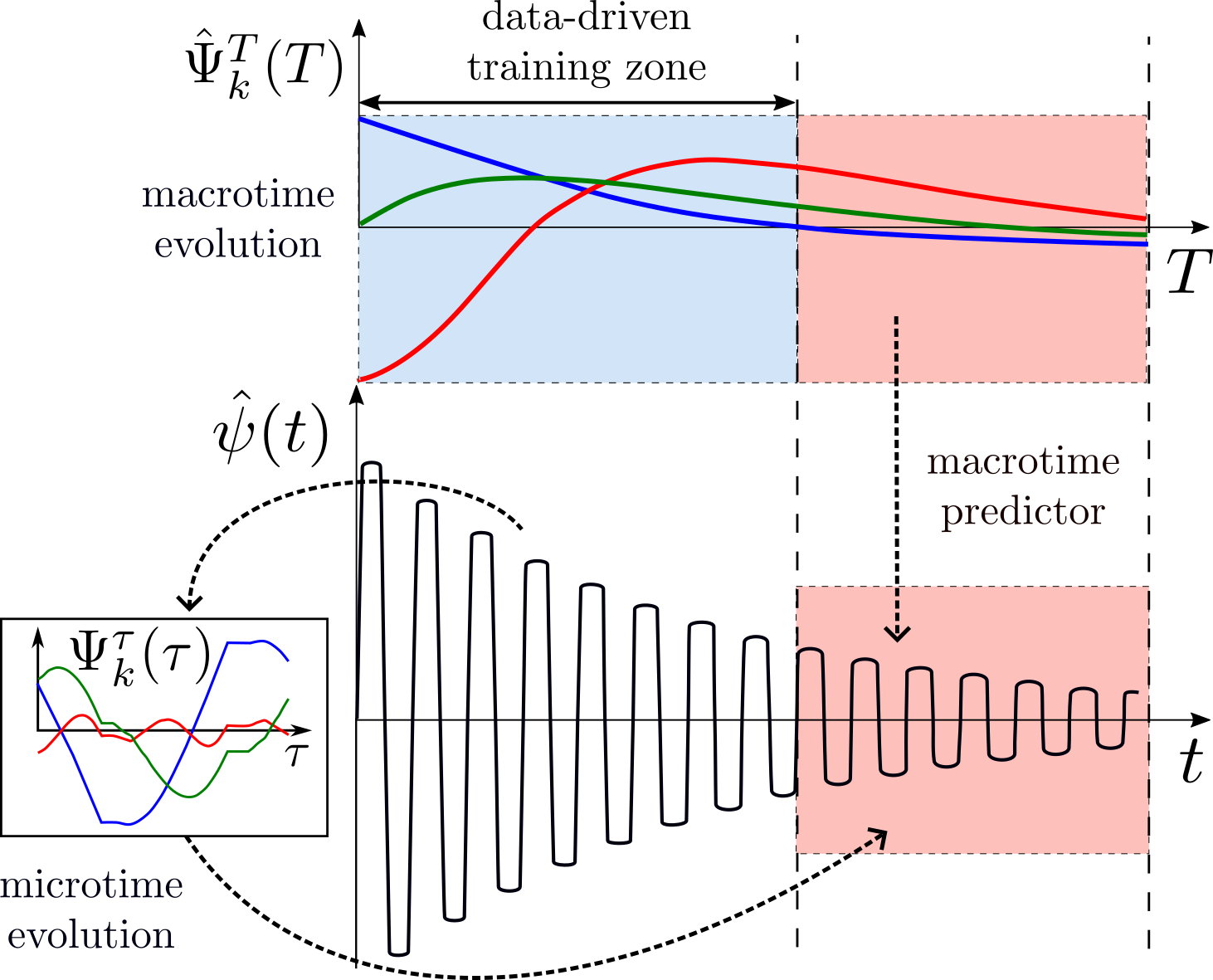}
		\caption{Data-driven macrotime integrator.}
		\label{fig:hodmd-macro}
	\end{figure}

	This is schematically illustrated in figure \ref{fig:hodmd-macro}, where $\psi$ denotes the nonlinear response $\boldsymbol{\varepsilon}^{p, l - 1}$ particularized in a spatial location.
	
    \subsection{Corrector}
    \label{subsec:corrector}
    The quality of the prediction \eqref{eq:3PGD-multiscale-predicted} should be compared with a full integration of the plasticity up to $T_N$, that is
	\begin{equation}
		\label{eq:history-dependent-full}
		\ctens{\varepsilon}^{p} =  \mathcal{N}(\ctens{\varepsilon}, \bar{\varepsilon}^p_{T_N}).
	\end{equation} 
	However, as already discussed, evaluations in \eqref{eq:history-dependent-full} are unfeasible when $T_N \gg T_K$ and when too many spatial nodes $N_{\vect{x}}$ are considered. Let us assume that this task, however, can be performed locally for a few reference spatial locations $\mathbf{x}_r = \{ \vect{x}_1^r, \dots, \vect{x}_J^r \}$, with $1 < J < N_{\vect{x}}$, like in sparse-sampling-based approaches (the locations can, for instance, be selected as the ones having the highest accumulated plastic strain $\bar{\varepsilon}^p_{T_K}$).
	
	In this sparse framework, instead of considering \eqref{eq:history-dependent-full}, the correction of $\hat{\ctens{\varepsilon}}^{p, (l - 1)}$ is based on employing its reduced counterpart over the set $\mathbf{x}_r$, which can be denoted as
	\begin{equation}
		\label{eq:history-dependent-hr}
		\ctens{\varepsilon}^{p}_{\mathbf{x}_r} =  \mathcal{N}_{\mathbf{x}_r}(\ctens{\varepsilon}, \bar{\varepsilon}^p_{T_N}).
	\end{equation}
	
	The predictor \eqref{eq:3PGD-multiscale-predicted} is corrected updating the macro modes by solving the following minimization problem:
	\begin{equation}
		\label{eq:minimization-update}
		\min_{\{\Delta \Psi^{T}_k\}_{k=1}^m} \norm{ 
			\sum_{k = 1}^m \Psi^{\vect{x}}_k(\vect{x}) \Psi^{\tau}_k(\tau) \left(\hat{\Psi}^T_k(T)  + \Delta \Psi^{T}_k(T)\right) - \ctens{\varepsilon}^p(\vect{x}, t) }_{\Omega_r \times \hat{I}},
	\end{equation}
	where $\norm{\bullet}_{\Omega_r \times \hat{I}} = \int_{\Omega_r} \int_{\hat{I}} \bullet \dd{\vect{x}} \dd{t}$ denotes a norm suitably defined over the reduced spatial domain $\Omega_r$ and the temporal prediction interval $\hat{I}$.
	
	Problem \eqref{eq:minimization-update} can be recasted in a weighted residual form, after having introduced suitable test functions $\{\Phi^{T}_k\}_{k=1}^m$, by
	\begin{small}
		\begin{equation}
			\label{eq:update-weighted-res}
			\int_{\Omega_r \times \hat{I}} \left( \sum_{k = 1}^m \Psi^{\vect{x}}_k(\vect{x}) \Psi^{\tau}_k(\tau) \Phi^{T}_k(T) \right) \left( \sum_{l = 1}^m \Psi^{\vect{x}}_l(\vect{x}) \Psi^{\tau}_l(\tau) \Delta \Psi^{T}_l(T) - \hat{e} (\vect{x}, t) \right) \dd{\vect{x}} \dd{t} = 0,
		\end{equation}
	\end{small}
	
	where $\hat{e} (\vect{x}, t)$ simply corresponds to the prediction error function, which can be expressed into a time-separated form after having rearranged $\ctens{\varepsilon}^p(\vect{x}, t)$ as $\ctens{\varepsilon}^p(\vect{x}, \tau, T)$:
	\begin{equation}
		\hat{e} (\vect{x}, \tau, T) = \sum_{k = 1}^m \Psi^{\vect{x}}_k(\vect{x}) \Psi^{\tau}_k(\tau) \hat{\Psi}^T_k(T) - \psi(\vect{x}, \tau, T).
	\end{equation}
	The following integrals can be defined, for all $k, l = 1, \dots, m$,
	\begin{equation}
		a_{kl} = \left( \int_{\Omega_r} \Psi^{\vect{x}}_k(\vect{x}) \Psi^{\vect{x}}_l(\vect{x}) \dd{\vect{x}} \right) \left( \int_{I^\tau} \Psi^{\tau}_k(\tau) \Psi^{\tau}_l(\tau) \dd{\tau} \right)
	\end{equation}
	and
	\begin{equation}
		b_k(T) =  \int_{I^\tau} \left( \int_{\Omega_r} \Psi^{\vect{x}}_k(\vect{x}) \hat{e} (\vect{x}, \tau, T) \dd{\vect{x}} \right) \Psi^{\tau}_k(\tau) \dd{\tau}.
	\end{equation}
	With these definitions made, equation \eqref{eq:update-weighted-res} can be rewritten as
	\begin{equation}
		\label{eq:update-weighted-res-simplified}
		\int_{\hat{I}^T} \sum_{k = 1}^m  \Phi^{T}_k(T) \sum_{l = 1}^m \Delta \Psi^{T}_l(T) a_{kl} \dd{T} = \int_{\hat{I}^T} \sum_{k = 1}^m  \Phi^{T}_k(T) b_k (T) \dd{T}.
	\end{equation}
	At this point, problem \eqref{eq:update-weighted-res-simplified} can be easily solved using finite elements in time, among other possibilities.
	
	In the above definitions, the time intervals $I^\tau$ and $\hat{I}^T$ are the ones associated to the micro and macro scales, respectively. In particular, the one related to the macroscale keeps the hat notation since it concerns the forecasting interval.

	\subsubsection{Enrichment}
	Once the optimal macrotime correction modes $\left\{\Delta \Psi^{T}_k(T)\right\}_{k=1}^m$ satisfying \eqref{eq:update-weighted-res-simplified} have been determined, a global enrichment step can be performed. This consists in adding ulterior modes $m^{\star} - m - 1$ to enrich the PGD approximation, solving the minimization problem
	\begin{equation}
		\label{eq:minimization-enrichment}
		\min_{\{ \Psi^{\vect{x}}_k, \Psi^{\tau}_k, \Psi^{T}_k\}_{k=m+1}^{m^\star}} \norm{ 
			\sum_{k = m+1}^{m^\star} \Psi^{\vect{x}}_k(\vect{x}) \Psi^{\tau}_k(\tau) \Psi^T_k(T) - \hat{e}^{\text{update}} (\vect{x}, \tau, T) }_{\Omega_r \times \hat{I}},
	\end{equation}
	where 
	\begin{equation}
		\hat{e}^{\text{update}} (\vect{x}, \tau, T) = \ctens{\varepsilon}^p(\vect{x}, \tau, T) - \sum_{k = 1}^m \Psi^{\vect{x}}_k(\vect{x}) \Psi^{\tau}_k(\tau) \left(\hat{\Psi}^T_k(T) + \Delta \Psi^{T}_k(T)\right).
	\end{equation}
	The minimization problem can be rewritten in the following weighted residual form 
	\begin{equation}
		\label{eq:weak-form-enrichment}
		\int_{\Omega_r \times \hat{I}} \sum_{k = m+1}^{m^\star} \Phi_k(\vect{x}, \tau, T) \left( \sum_{k = m+1}^{m^\star} \Psi^{\vect{x}}_k(\vect{x}) \Psi^{\tau}_k(\tau) \Psi^T_k(T) - \hat{e}^{\text{update}} (\vect{x}, \tau, T) \right) = 0.
	\end{equation}
	In problem \eqref{eq:weak-form-enrichment} the following test function has been introduced
	\begin{equation}
		\Phi_k(\vect{x}, \tau, T) =  \Phi^{\vect{x}}_k(\vect{x}) \Psi^{\tau}_k(\tau) 
		\Psi^T_k(T) + \Psi^{\vect{x}}_k(\vect{x}) \Phi^{\tau}_k(\tau) 
		\Psi^T_k(T) + \Psi^{\vect{x}}_k(\vect{x}) \Psi^{\tau}_k(\tau) 
		\Phi^{T}_k(T),
	\end{equation}
	where $\Phi^{\vect{x}}_k, \Phi^{\tau}_k$ and $\Phi^{T}_k$ are three independent test functions, for the space, micro time and macrotime problems, respectively. Finally, the solution of \eqref{eq:weak-form-enrichment} is obtained by means of a fixed-point alternating direction strategy, as usual in PGD-based procedures \cite{PGD,PGD-1}.
	
	The corrected (optimal) predictor, after the update-enrichment procedure is then defined as
	\begin{equation}
		\hat{\ctens{\varepsilon}}^{p, \star}(\vect{x}, t) =
		\underbrset{\text{update}}{\sum_{k = 1}^m \Psi^{\vect{x}}_k(\vect{x}) \Psi^{\tau}_k(\tau) \left(\hat{\Psi}^T_k(T) + \Delta \Psi^{T}_k(T)\right)} + \underbrset{\text{enrichment}}{\sum_{k = m+1}^{m^\star} \Psi^{\vect{x}}_k(\vect{x}) \Psi^{\tau}_k(\tau) \Psi^T_k(T)}.
	\end{equation}

   \subsection{Summary of the solving scheme}
   
\begin{figure}[H]
   	\centering
   	\includegraphics[width=0.8\textwidth]{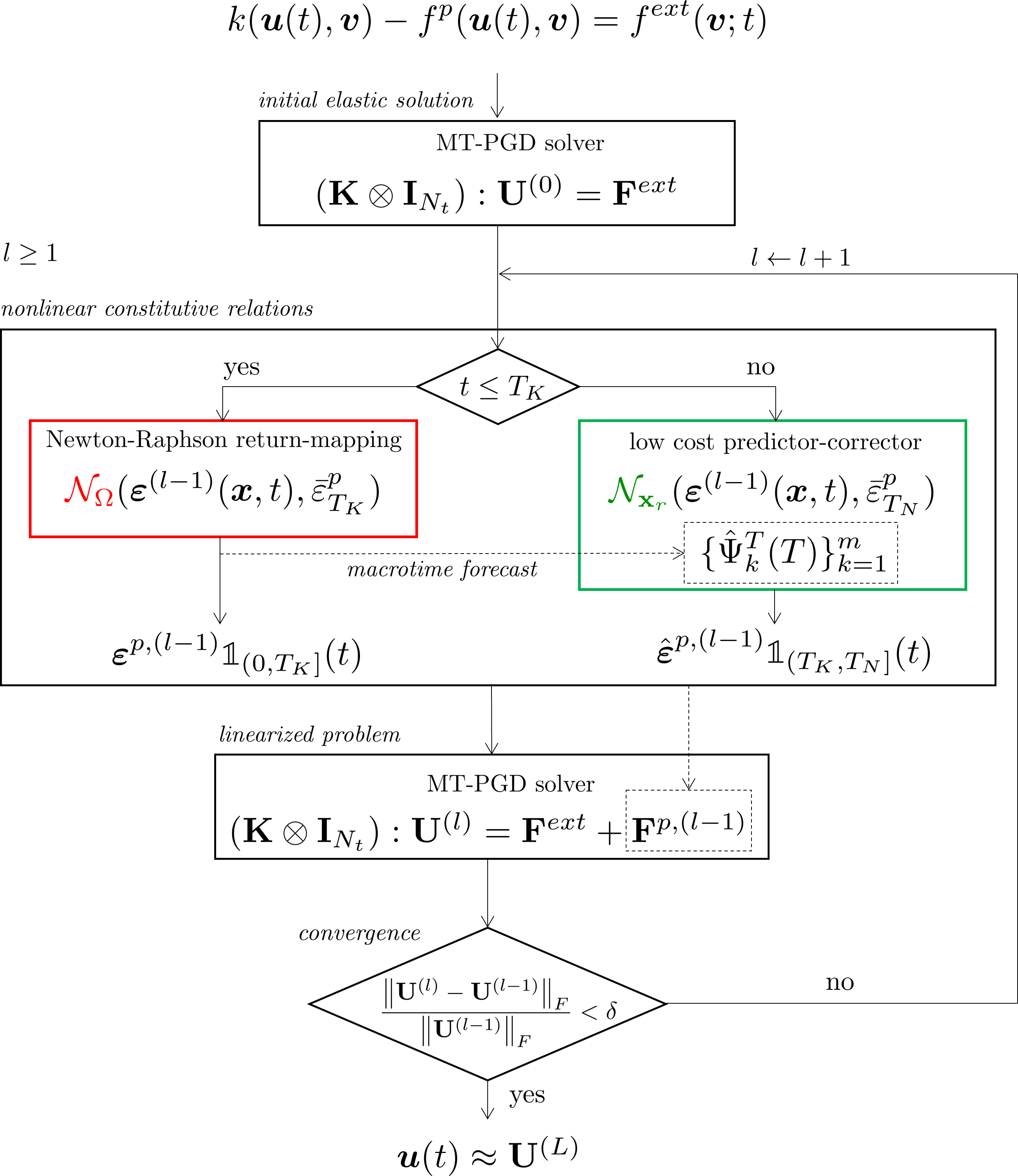}
   	\caption{Data-driven MT-PGD solving scheme for elasto-plasticity.}
   	\label{fig:solving-scheme}
   \end{figure}

   The overall solving procedure is summarized in the flowchart in figure \ref{fig:solving-scheme}.
   
   When $t \in (0, T_K]$, the algorithm consists of computing the quasi-static elasto-plastic response, using the PGD-based approach from \cite{pgd-multiscale-3}. 
   
   When, $t \in (T_K, T_N]$, a snapshot of the plastic strain tensor $\vect{\varepsilon}^p\mathbb{1}_{(0, T_K]}(t)$ is exploited to build a data-driven forecasting model of the nonlinear constitutive relations. The prediction $\hat{\vect{\varepsilon}}^p\mathbb{1}_{(T_K, T_N]}(t)$ is, then, corrected by means of a sparse selection of reference spatial locations $\mathbf{x}_r$. Once the corrected (optimal) prediction $\hat{\vect{\varepsilon}}^{p, \star}\mathbb{1}_{(T_K, T_N]}(t)$ is available, the linearized problem is assembled up to $T_N$ and, finally, efficiently solved via the MT-PGD.
   
    \section{Results and discussion}
    \label{sec:application-case}
    
    This section presents the numerical results over the same test-case considered in \cite{pgd-multiscale-3}, consisting of a uniaxial load-unload tensile test over a dog-bone shaped steel specimen. The loading in a Dirichlet datum $u_D(t)$ having constant amplitude applied to both sides of the specimen.
    
    The material has a Young modulus $E = 210$ GPa and a Poisson ratio $\nu = 0.3$ and its plasticity law (linear isotropic hardening) is characterized by an initial yield stress $\sigma_{y, 0} = 205$ MPa and a linear hardening coefficient $H = 2$ GPa. The imposed displacement has a maximum amplitude $u_D^{max} = 0.125$ mm and a single cycle (load-unload-load) time has duration $T_1 = 4 u_D^{max}/v_l = 20$ s, where $v_l = 0.025$ mm/s is the load rate ensuring a quasi-statics simulation.
    
\begin{figure}[H]
    	\centering
    	\begin{subfigure}{0.45\textwidth}
    		\centering
    		\includegraphics[width=1\textwidth]{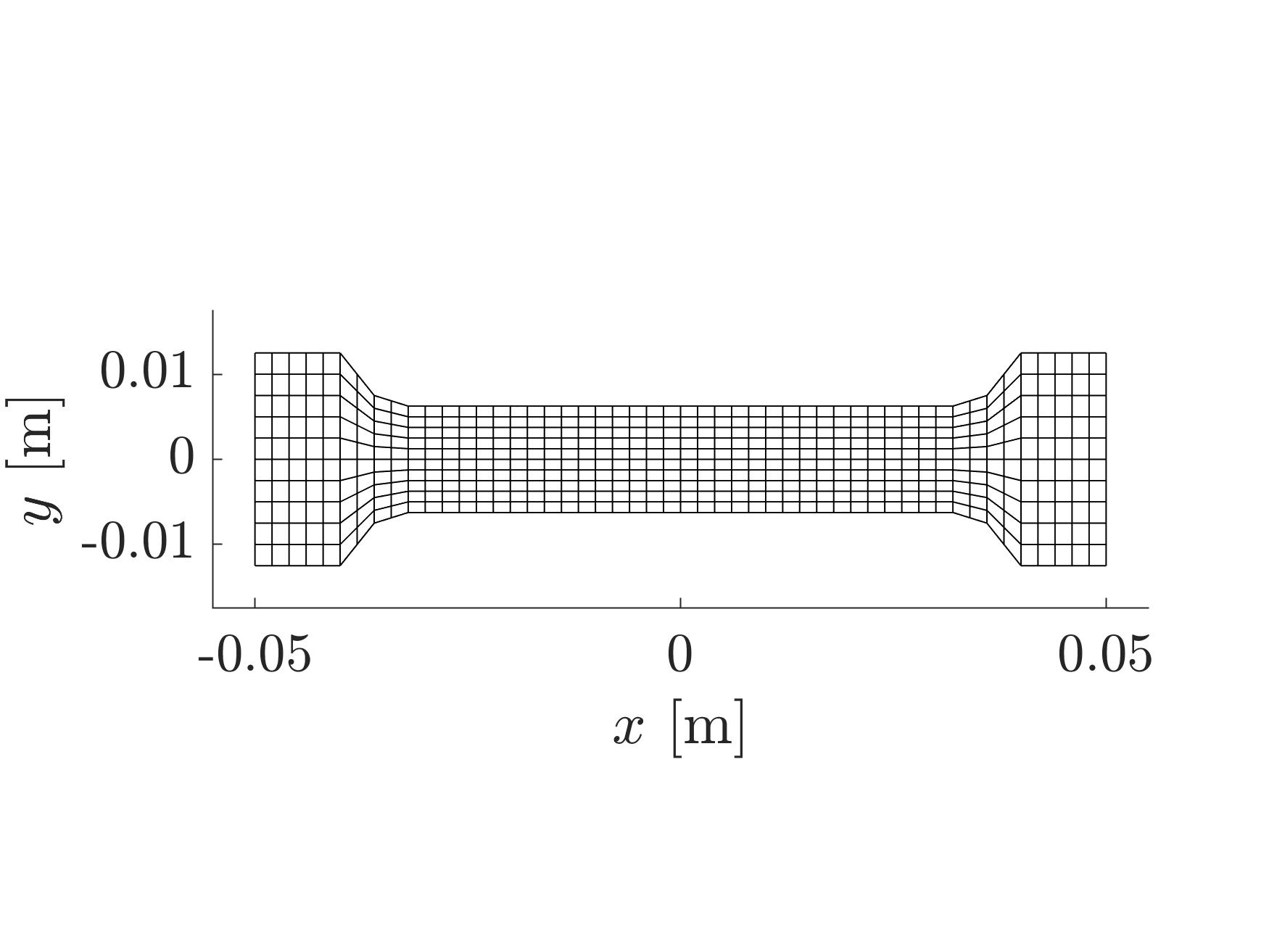}
    	\end{subfigure}
    	\begin{subfigure}{0.45\textwidth}
    		\centering
    		\includegraphics[width=1\textwidth]{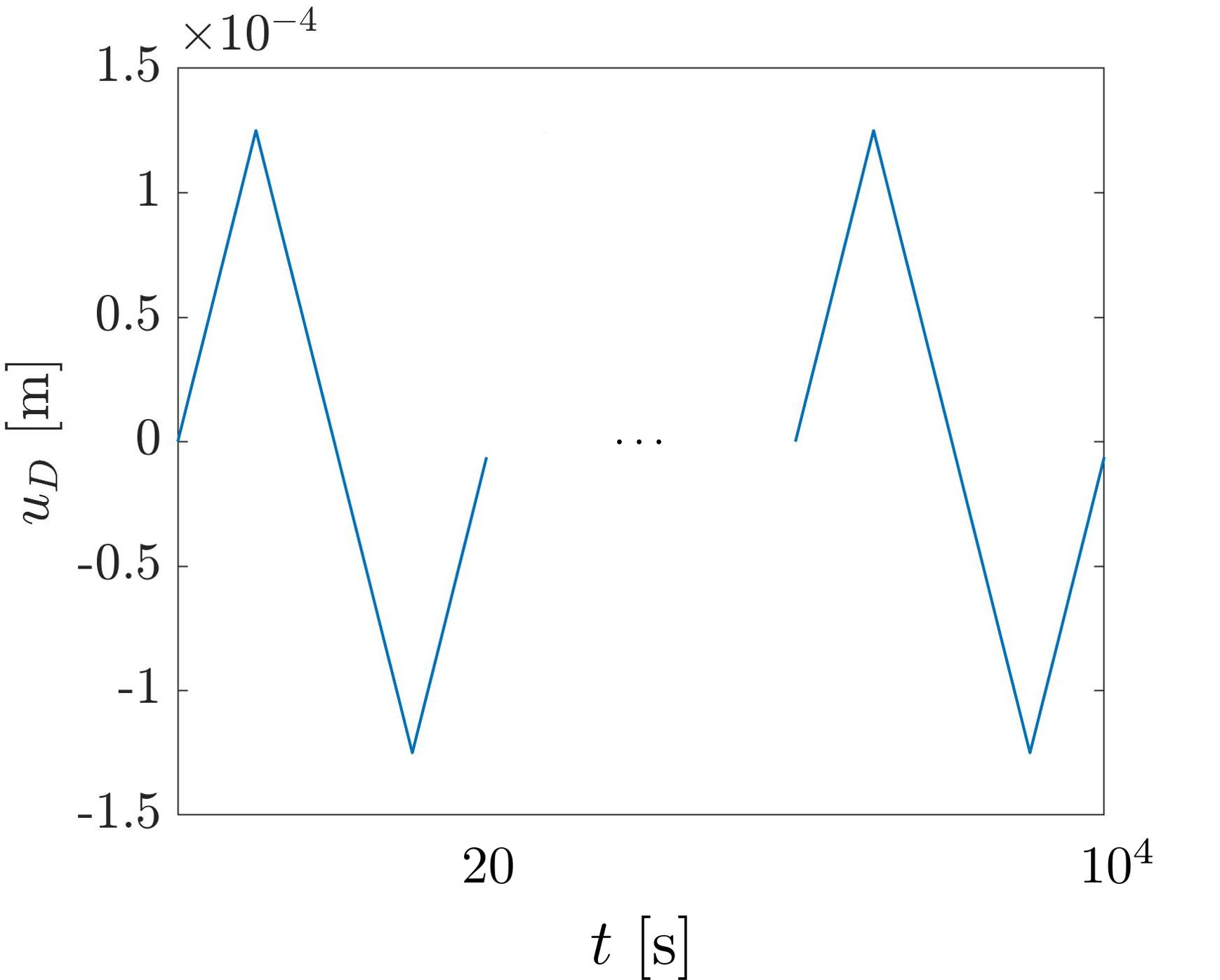}
    	\end{subfigure}
    	\caption{Discretized geometry (left) and imposed displacement (right).}
    	\label{fig:test-case}
    \end{figure}

    Figure \ref{fig:test-case} shows the two-dimensional discretized geometry, consisting of $N_e = 500$ quadrilateral elements and $N_x = 561$ mesh nodes. The cyclic loading is also shown in figure \ref{fig:test-case}, where the red line represents the average. The simulation is performed using the algorithms from \cite{pgd-multiscale-3} up to $K = 500$ cycles. The simulation is then extended in almost real-time to $N = 1500$ cycles using the data-driven modeling of the nonlinear term. Let us denote with $T_K$ and $T_N$ the ending times of the cycles $K$ and $N$, respectively. The time intervals $I_K = (0, T_K]$ and $I_N = (T_K, T_N]$ are both discretized in equispaced $N_t^{(K)} = N_t^{(N)} = 4\cdot 10^5$ time instants.
    
\begin{figure}[H]
    	\centering
    	\begin{subfigure}{0.45\textwidth}
    		\centering
    		\includegraphics[trim={0 300 0 300},clip,width=1\textwidth]{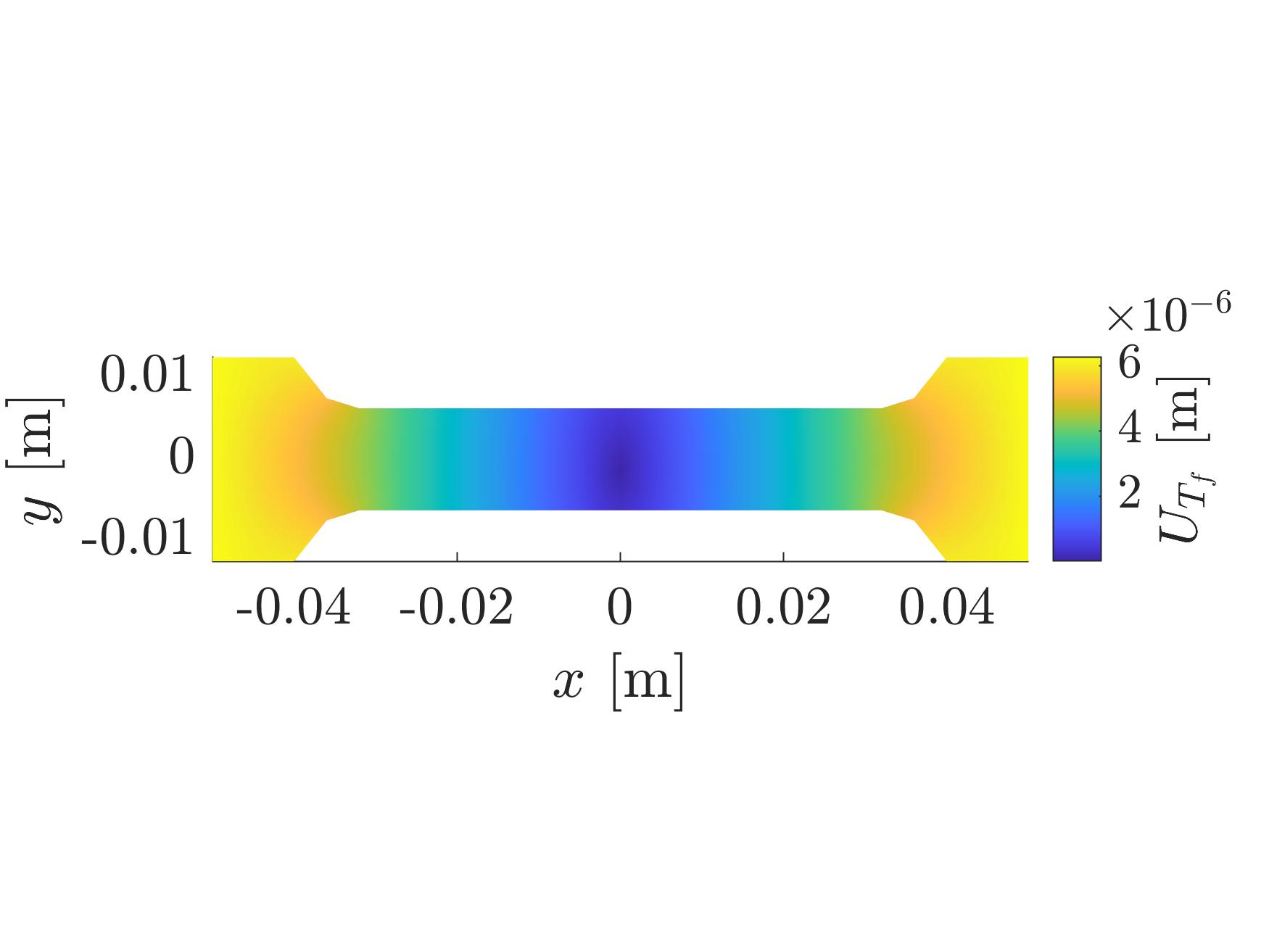}
    	\end{subfigure}
    	\begin{subfigure}{0.45\textwidth}
    		\centering
    		\includegraphics[trim={0 300 0 300},clip,width=1\textwidth]{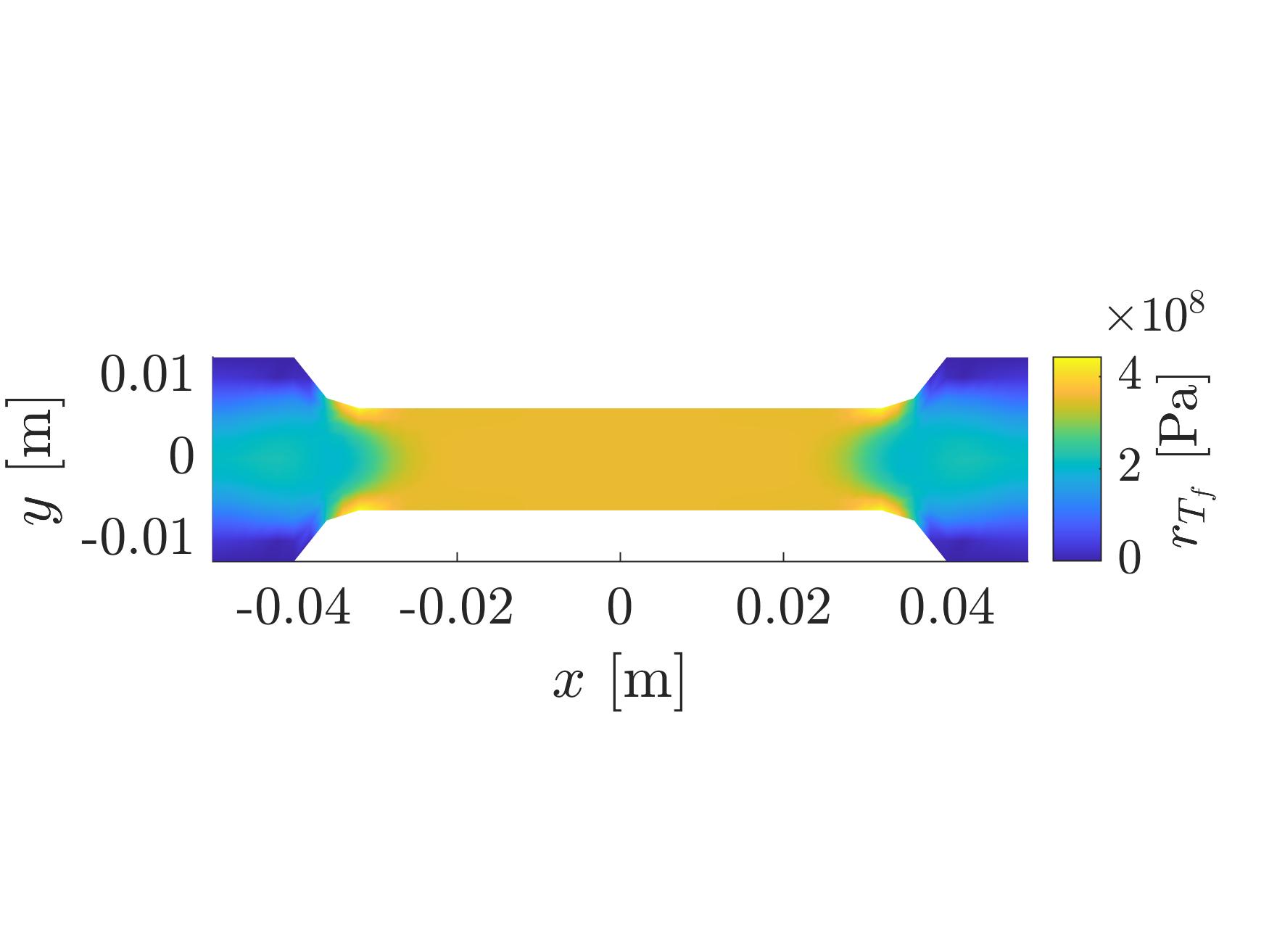}
    	\end{subfigure}
    	\caption{Displacement field (left) and isotropic hardening (right) where $T_f = T_K$.}
    	\label{fig:test-case-results}
    \end{figure}

    Figure \ref{fig:test-case-results} gives the magnitude of the displacement field and the isotropic hardening function computed at the time $T_K = 10^4$ s.
    
    The plastic strain tensor history $\ctens{\varepsilon}^{p} \mathbb{1}_{I_K} (t)$ is here used to build the data-driven model as described in section \ref{sec:dd-model}. The related snapshot is defined as
	\begin{equation}
		\label{eq:snapshot-full}
		\SDtens{\Psi} = \begin{pmatrix}
			\SDvect{\uppsi}_1 | & \cdots & | 	\SDvect{\uppsi}_{N_t^{(K)}}
		\end{pmatrix} \in \bbR^{3 N_\vect{x} \times N_t^{(K)}}
	\end{equation}
	where $\SDvect{\uppsi}_j$ is a column vector containing the numerical approximations of $\ctens{\varepsilon}^{p}(\vect{x}, t_j)$ in all the $N_\vect{x}$ spatial mesh points, for $j = 1, \dots, N_t^{(K)}$. The column vectors account for the concatenation of the three components $\ctens{\varepsilon}^p = (\varepsilon^p_{11}, \varepsilon^p_{12}, \varepsilon^p_{22})$ for the two-dimensional case here analyzed.
	
	Figure \ref{fig:test-case-results} gives the magnitude of the plastic strain computed at final time $T_K$.
	
	\begin{figure}[H]
		\centering
		\includegraphics[width=0.55\textwidth]{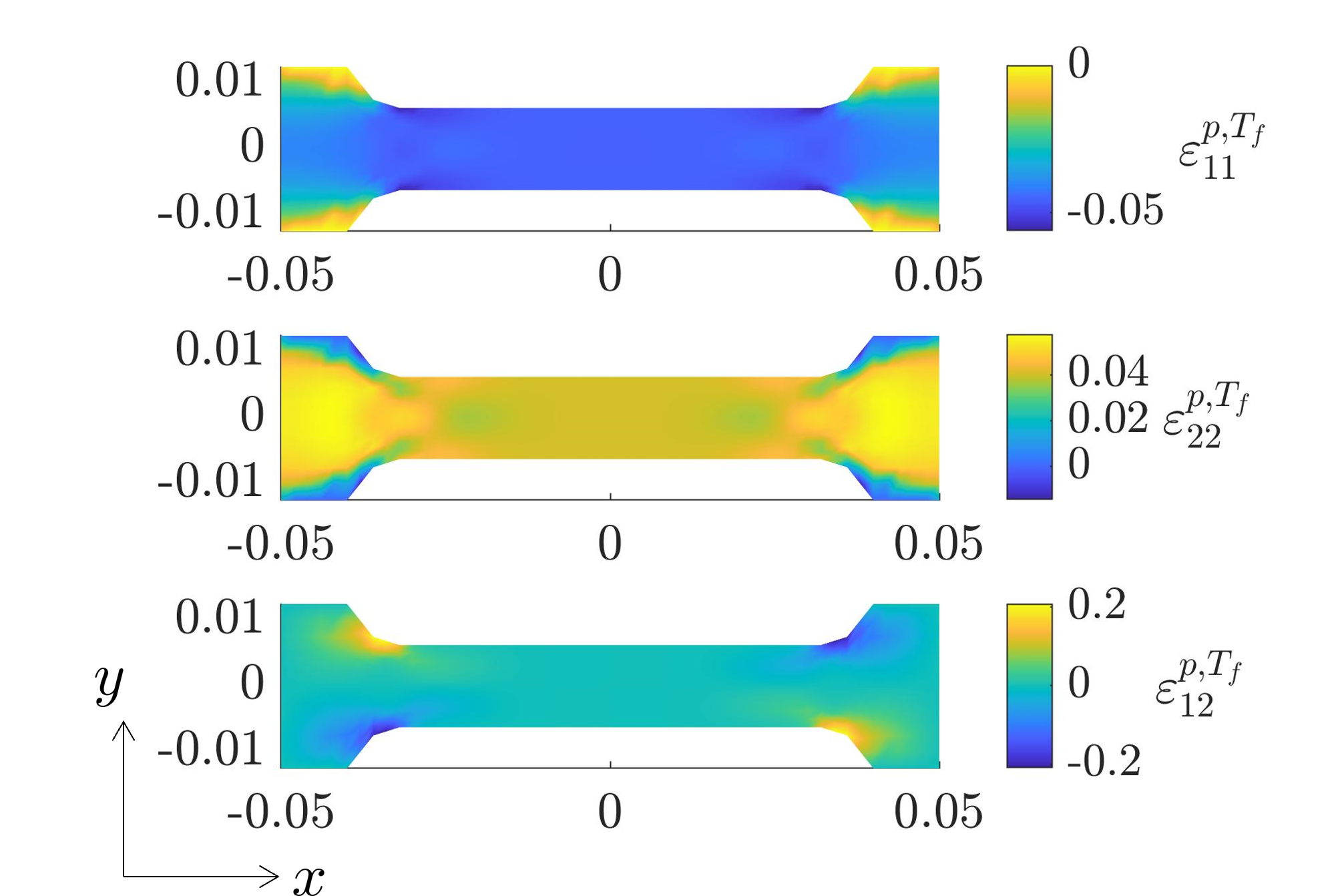}
		\caption{Plastic strain components where $T_f = T_K$.}
		\label{fig:plastic-strain-results}
	\end{figure}

	To quickly illustrate the time evolution of $\ctens{\varepsilon}^{p} (\vect{x}, t)$, a POD-based reduced representation \cite{POD,MOR-2} of the snapshot \eqref{eq:snapshot-full} can be considered, being the approximation
	\begin{equation}
		\label{eq:POD-approx}
		\ctens{\varepsilon}^{p}(\vect{x}, t) \approx \ctens{\varepsilon}^{p,\text{POD}}(\vect{x}, t) = \sum_{k = 1}^m w_k^{\vect{x}}(\vect{x}) \alpha_k^t(t),
	\end{equation}
	where the functions $w_k^{\vect{x}}(\vect{x})$ and $\alpha_k^t(t)$, $k = 1, \dots, m$ are the space and time modes.
	
\begin{figure}[H]
		\centering
		\includegraphics[width=0.6\textwidth]{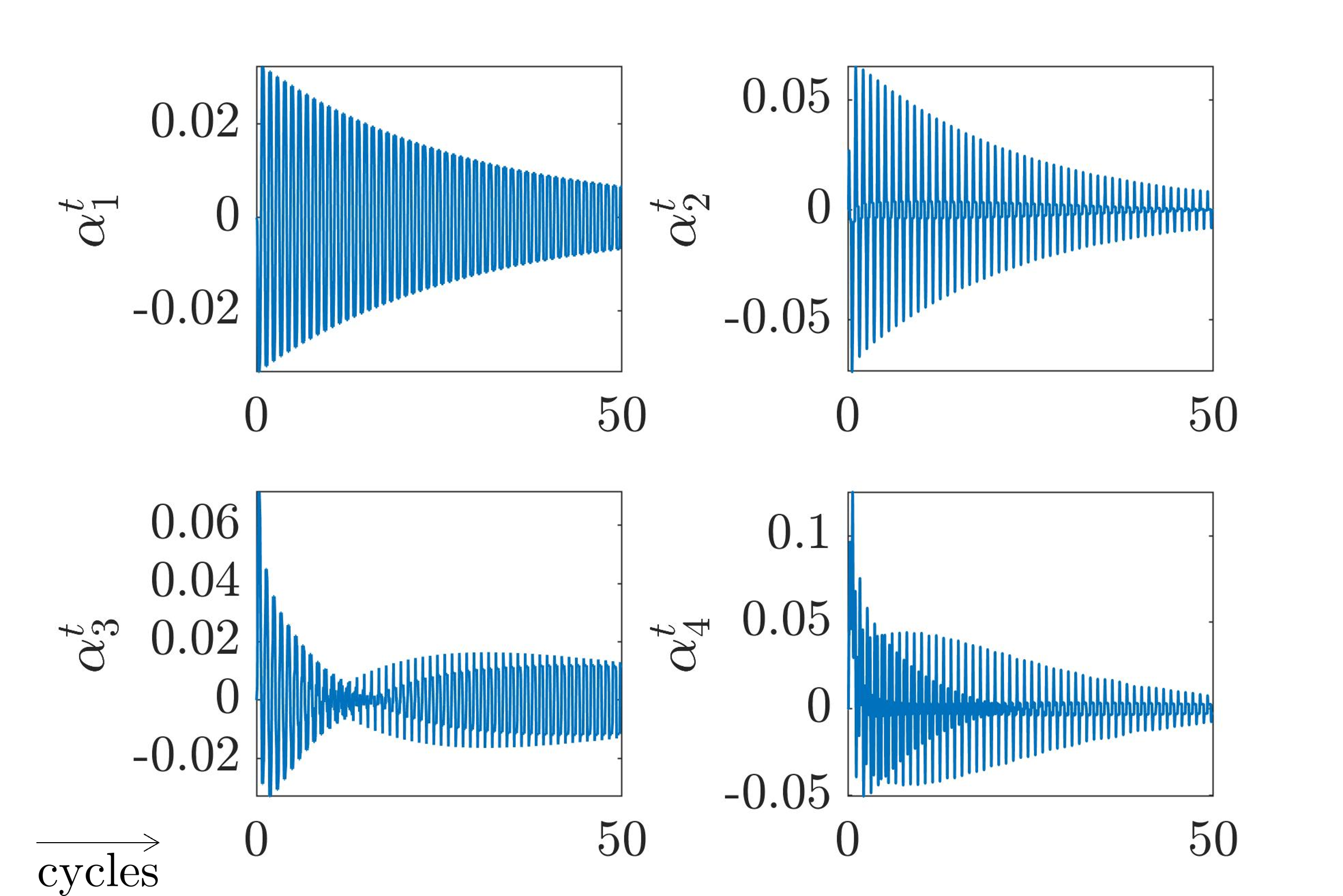}
		\caption{First four normalized POD time modes of $\psi (\vect{x}, t)$.}
		\label{fig:plastic-strain-POD-time-modes}
	\end{figure}

\begin{figure}[H]
		\centering
		\includegraphics[width=0.85\textwidth]{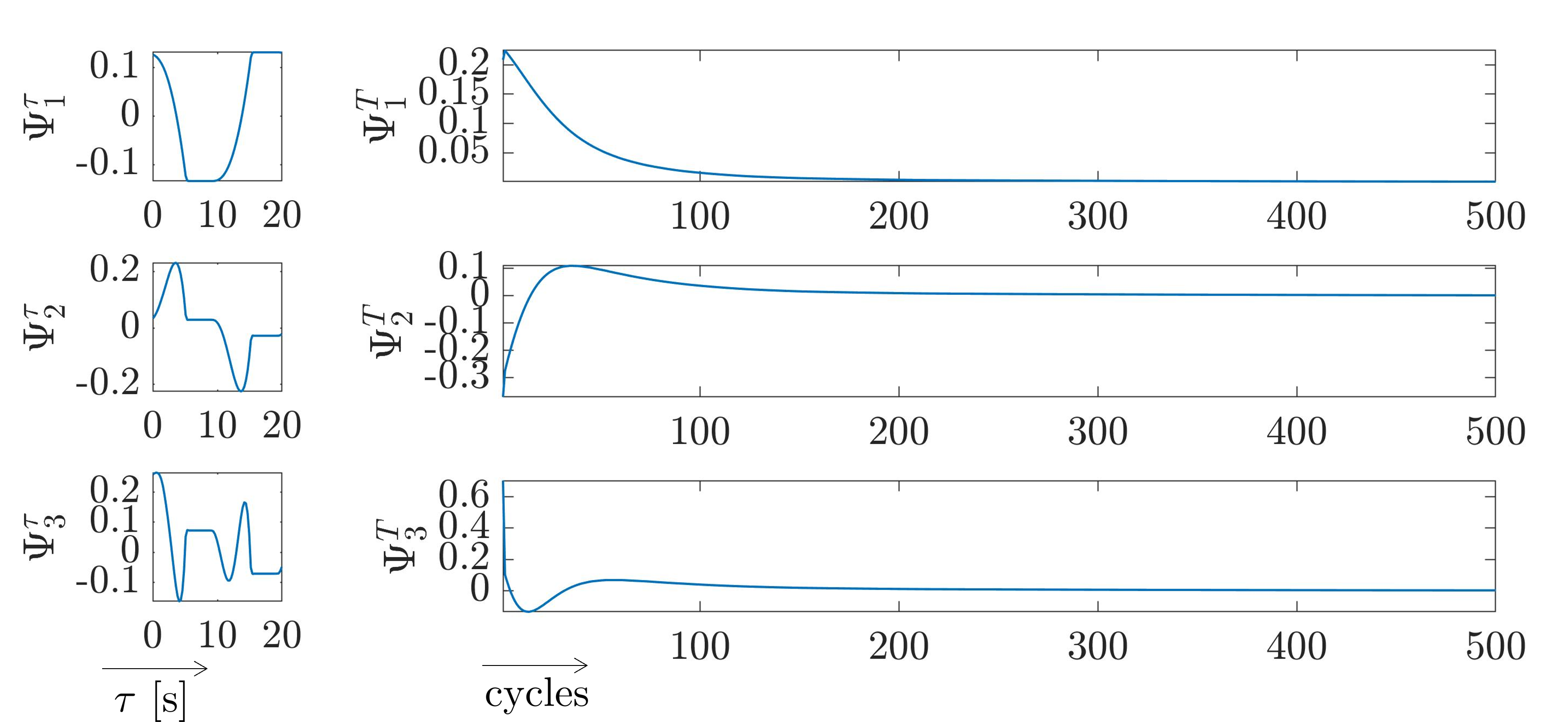}
		\caption{First three normalized MT-PGD time modes of $\psi (\vect{x}, t)$.}
		\label{fig:plastic-strain-MTPGD-time-modes}
	\end{figure}

\begin{figure}[H]
		\centering
		\includegraphics[width=0.55\textwidth]{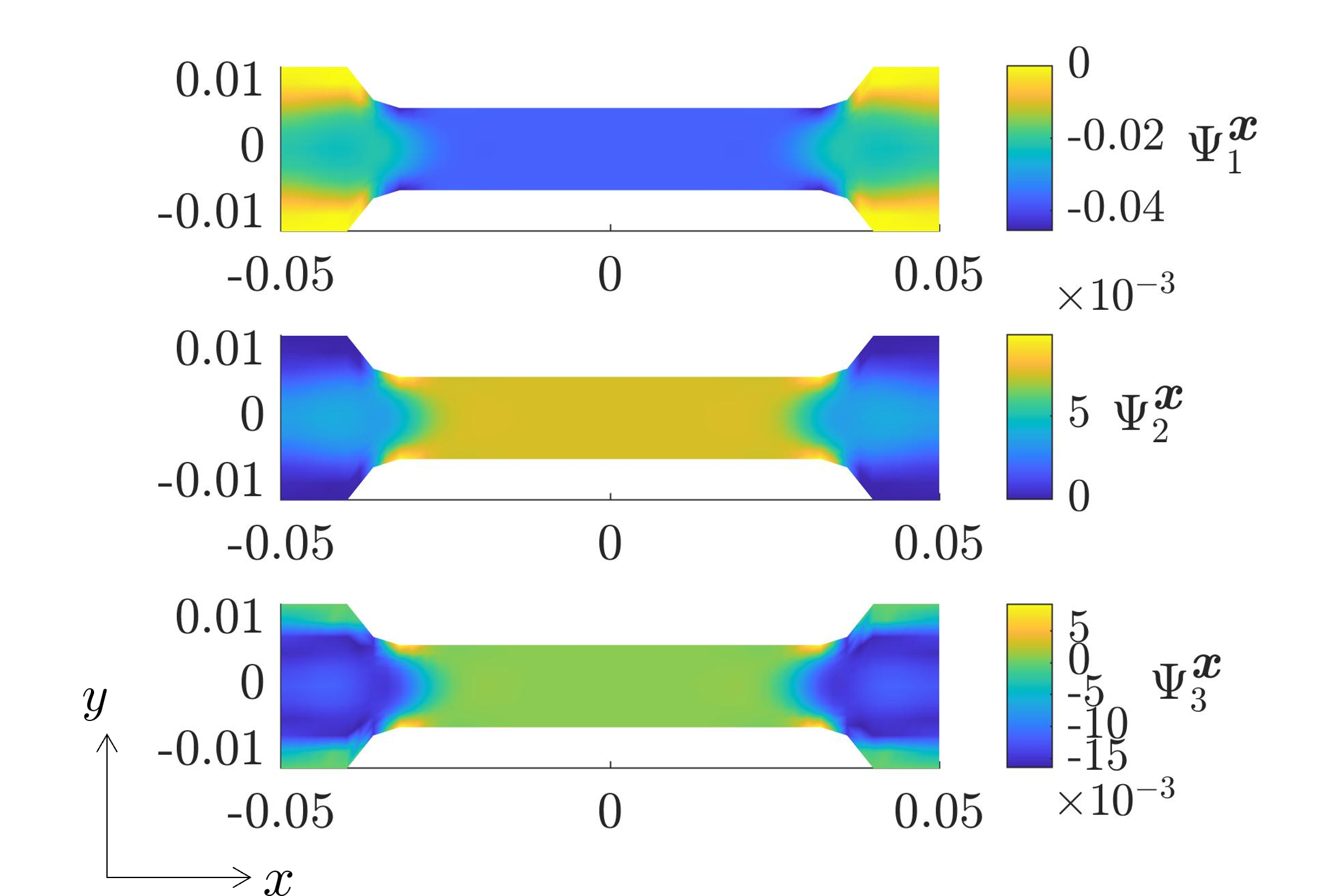}
		\caption{First three normalized MT-PGD space modes of $\psi (\vect{x}, t)$.}
		\label{fig:plastic-strain-MTPGD-space-modes}
	\end{figure}

	For instance, figure \ref{fig:plastic-strain-POD-time-modes} depicts the POD time functions over the first 50 cycles. Even though the functions exhibit a decay towards 0, the highly nonlinear patterns (at the cycle level) make difficult the construction of a prediction model able to track accurately the fast scale.
	The forecasting task is clearly simplified when considering the multi-time PGD approximation. In fact, figure \ref{fig:plastic-strain-MTPGD-time-modes} shows the micro time and macro time functions over the first $K = 500$ cycles. For the sake of brevity only the spatial modes related to the component $\varepsilon^p_{11}$ are shown in figure \ref{fig:plastic-strain-MTPGD-space-modes}. 
	
\begin{figure}[H]
		\centering
		\includegraphics[width=0.55\textwidth]{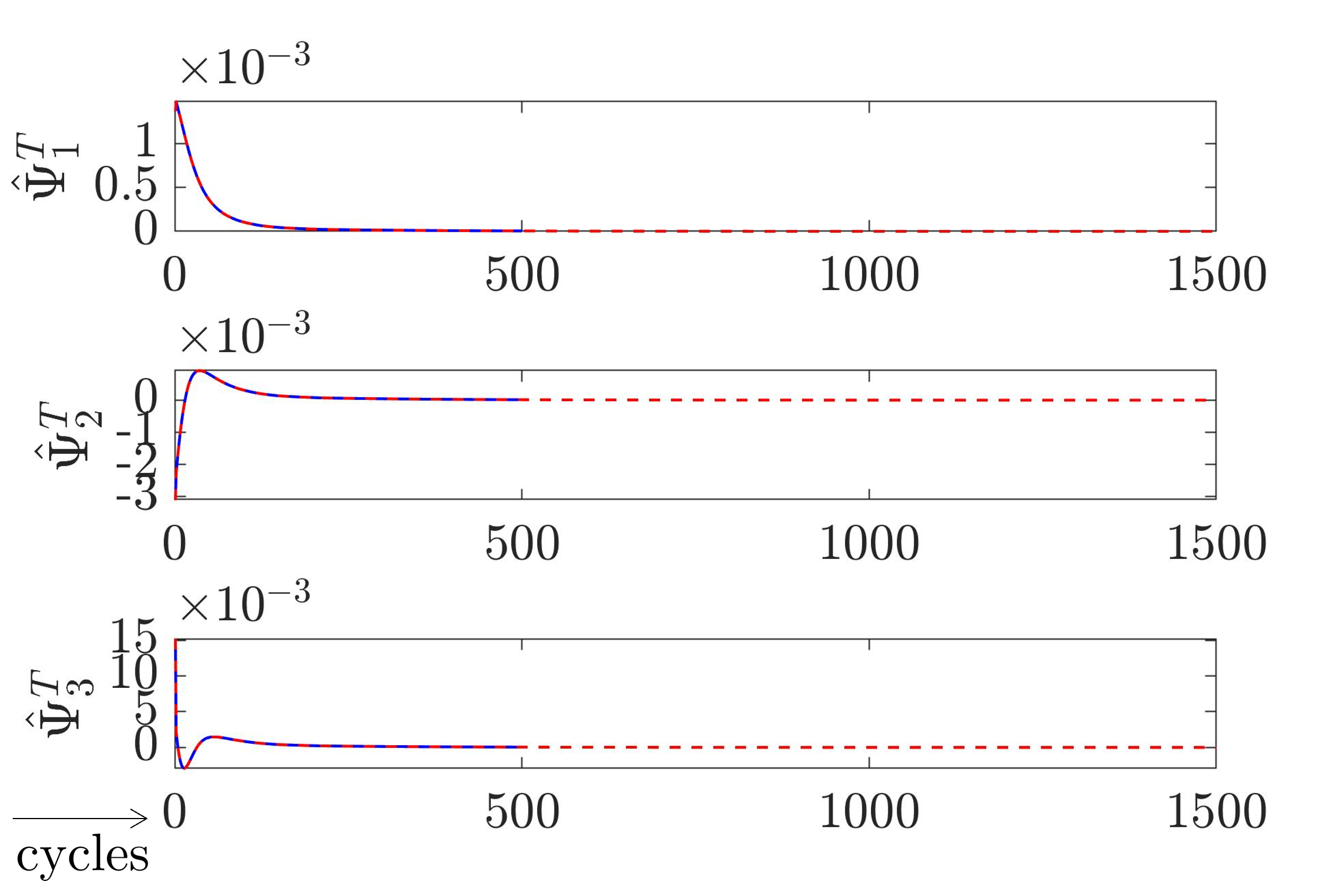}
		\caption{HODMD-based prediction of the macrotime modes.}
		\label{fig:plastic-strain-time-modes-hodmd}
	\end{figure}

	The HODMD-based extensions of the macrotime functions are shown in figure \ref{fig:plastic-strain-time-modes-hodmd} and used to predict the nonlinear response via \eqref{eq:3PGD-multiscale-predicted}.
	
	Letting $\norm{\bullet}_{\Omega \times \hat I} \int_{\Omega} \int_{\hat{I}} \bullet \dd{\vect{x}} \dd{t}$, the prediction error can be measured as
	\begin{equation}
		\hat{\epsilon} = \frac{\norm{\hat{\ctens{\varepsilon}}^{p}(\vect{x}, t) - \ctens{\varepsilon}^{p}(\vect{x}, t)}_{\Omega \times \hat I}}{\norm{\ctens{\varepsilon}^{p}(\vect{x}, t)}_{\Omega \times \hat I}},
	\end{equation}
	and amounts to $\hat{\epsilon} = 0.2146$. This discrepancy is recovered when through the correction step.
	In this case, 8 elements (more could be selected if needed) are enough to reduce the error to $\hat{\epsilon}^\star = 0.0109$. The elements have been selected as those having maximum effective plastic strain $\bar{\varepsilon}^p_{T_K}$.
	
\begin{figure}[H]
		\centering
		\includegraphics[width=0.95\textwidth]{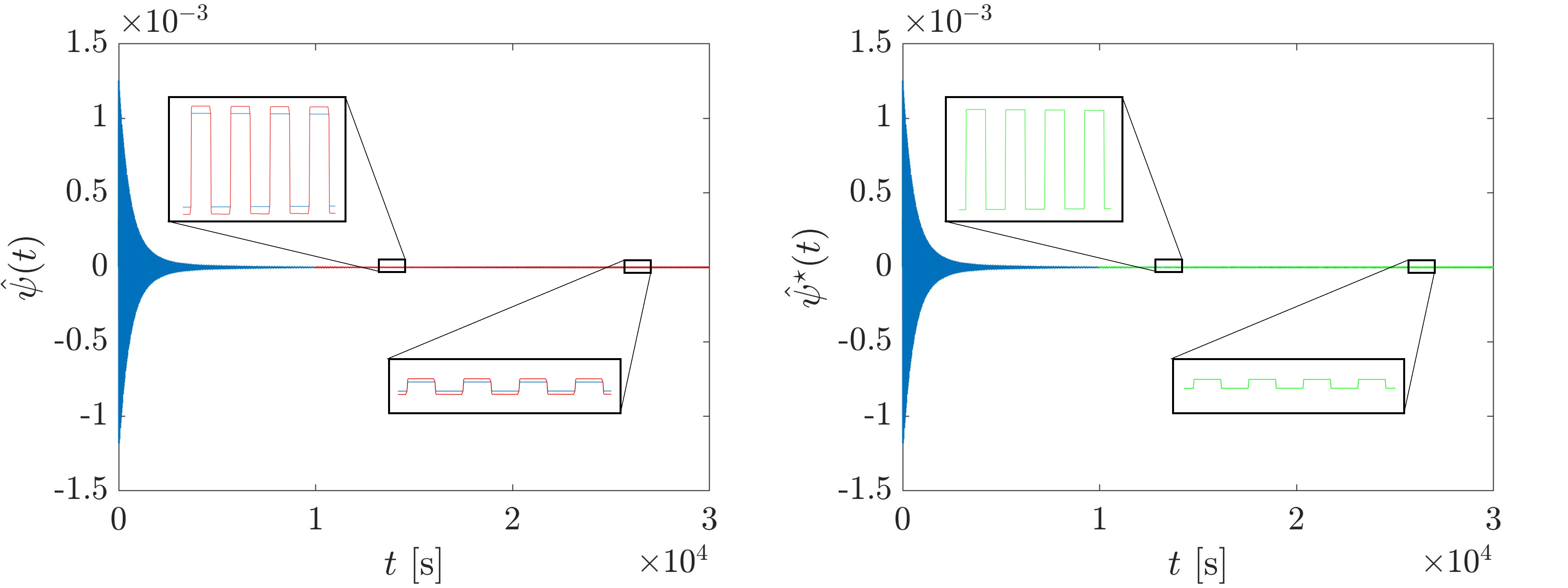}
		\caption{Reference (blue), predicted (red) and predicted-corrected (green) response.}
		\label{fig:signal-extrap}
	\end{figure}

\begin{figure}[H]
		\centering
		\includegraphics[width=0.5\textwidth]{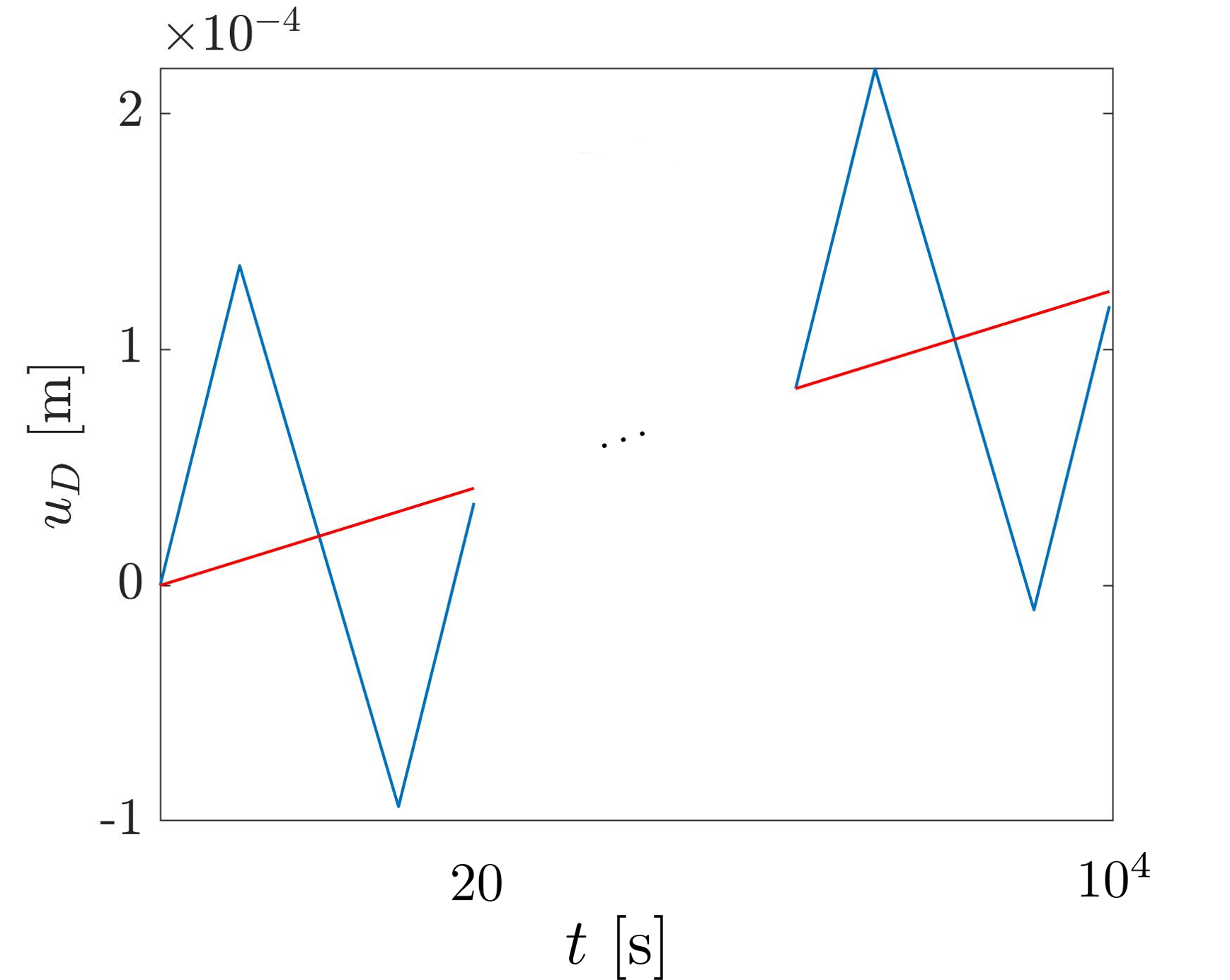}
		\caption{Cyclic displacement with linearly increasing average.}
		\label{fig:lin-disp}
	\end{figure}

\begin{figure}[H]
		\includegraphics[width=1\textwidth]{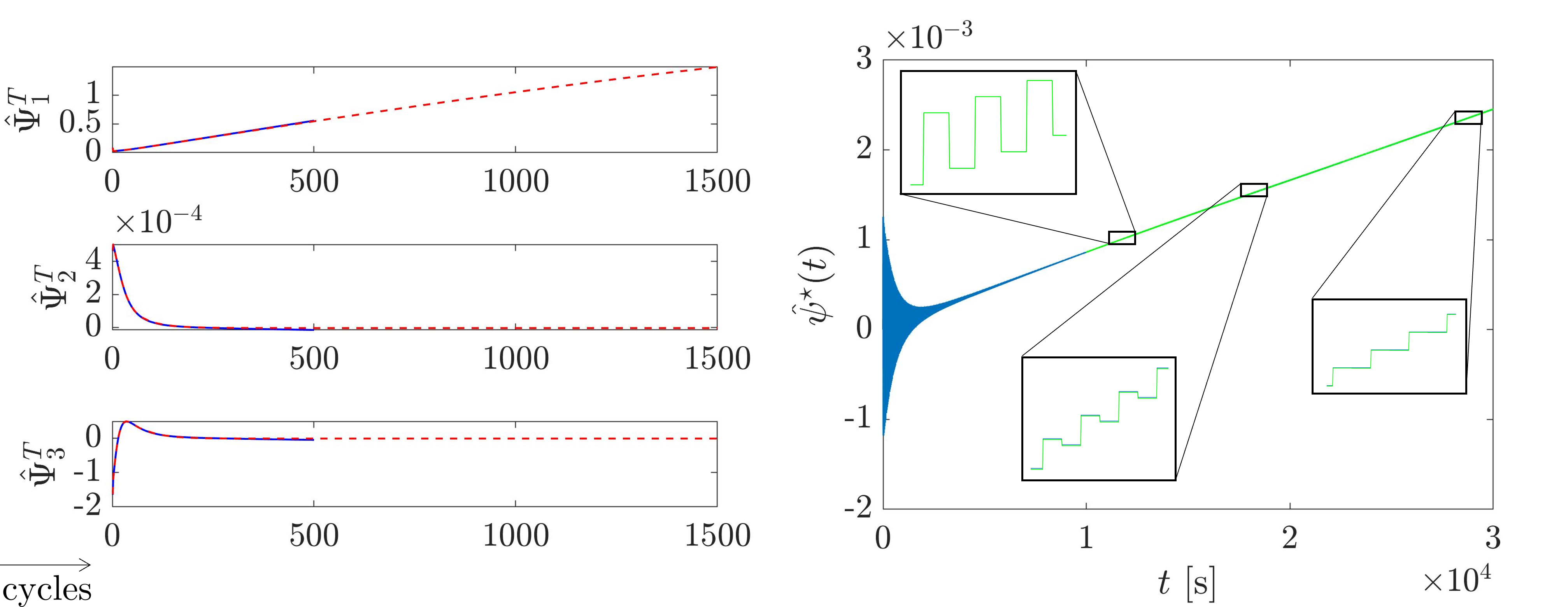}
		\caption{Macrotime modes predictions (left) and predicted-corrected response (right).}
		\label{fig:lin-disp-results}
	\end{figure}

	For a better understanding of the forecasting results, let us consider the time response $\psi(t) = \varepsilon_{11}^p(\vect{x}_c, t)$ in the center of the specimen $\vect{x}_c = (0, 0)$, illustrated in figure \ref{fig:signal-extrap}. Particularly, the loss of amplitude and the slightly inaccurate patterns of the predictor $\hat \psi$ are perfectly recovered by its corrected counterpart $\hat \psi^\star$. 
	The same procedure applies to other benchmark cases, such as the one of imposed displacement with linearly increasing average, depicted in figure \ref{fig:lin-disp}. The red line represents the average, whose slope is $u_D^{max}/T_f$ [m/s].
	Figure \ref{fig:lin-disp-results} shows the HODMD-based extension of the macrotime modes and the reconstructed signal at the center of the specimen $\vect{x}_c = (0, 0)$, whose evolving patterns are accurately captured.

	In terms of computational time gains, the performed tests show that the data-driven based evaluation of the nonlinear term (red versus green box in figure \ref{fig:solving-scheme}) has a speed-up factor of 2.3, approximately. Moreover, the overall solver time comparisons (figure \ref{fig:simulation-up-to-K} versus figure \ref{fig:solving-scheme}) shows a speed-up of 2.9, approximately. The additional gain around 0.6 comes from the separated space-microtime-macrotime format of the predicted right-hand-side. Indeed, as discussed in the introduction, the PGD solver assembly becomes faster when all the terms in the equation have separated representations.
	
    \section{Conclusions}
    \label{sec:conclusion}
	
	This work aims at reducing the computational complexity of numerical simulations in cyclic loading analyses, in particular when history-dependent nonlinear behaviors are considered. To this purpose, a novel time multiscale based data-driven modeling of the nonlinearity is proposed. The procedure makes use of multi-time PGD-based representations to separate the fast (micro) and slow (macro) time dynamics. 
		
	The first step consists in collecting the plastic strain history (and other nonlinear variables evolution, eventually) up to a given number of training cycles. Afterwards, the multi-time PGD \cite{pgd-multiscale-3} is used to decompose the time evolution in a multiscale manner, enabling the definition of a time integrator for the macrotime functions. Among other possible choices \cite{DMD-time-series}, the higher-order DMD is here used for the forecasting. Once the predictor of the nonlinear term is established, it is corrected by a few high-fidelity integrations of the plasticity up to the desired final time. The linearized problem is then solved efficiently using again the multi-time PGD. 
	
	Here below a brief recap of the computational and memory savings is given. For the sake of simplicity, the discussion considers the macroscale tracking all the cycles and the microscale evolving within a single cycle. 
	
	An incremental finite element based simulation based on $N_t^C$ increments for single cycle and considering $N$ cycles (thus $N_t^CN$ increments) would require an asymptotic complexity scaling as $\mathcal{O}(N_{\vect{x}}N_t^CN)$, where $N_{\vect{x}}$ is the number of spatial mesh points. The proposed procedure requires, a complexity of $\mathcal{O}(N_{\vect{x}}N_t^CK)$, with $K \ll N$ the number of training cycles, followed by a constant negligible complexity of the HODMD-based predictor. Afterwards, the simulation extension to $N$ cycles requires (a) the correction of the predictor based on the full-history integration over the reduced set of locations $\mathbf{x}_r = \{ \vect{x}_1^r, \dots, \vect{x}_J^r \}$, with $J \ll N_{\vect{x}}$, having complexity $\mathcal{O}(JN_t^CN)$, (b) the solution of the linearized problem employing the time multiscale PGD, with a complexity of $\mathcal{O}(N + N_t^C + N_{\vect{x}})$. This implies interesting computational gains observing the ratio $\frac{ \mathcal{O}(JN_t^CN) +  \mathcal{O}(N + N_t^C + N_{\vect{x}}) }{ \mathcal{O}(N_{\vect{x}}N_t^CN)}$, with $J \ll N_{\vect{x}}$. 
	
	It is worth noticing the advantages in terms of storage requirements in the final linearized problem. A usual time marching scheme requires the storage of time functions discretized in $N_t^CN$ points (where $N_t^C$ can be really high when a small step is required). Contrarily, when employing the multi-time PGD, the functions are stored as $N_t^C$ and $N$ points, for the microscale and macroscale, respectively. Moreover, thanks to their slow evolution, the macro functions can be reconstructed only by means of a few coefficients $p$. In this case, in terms of storage one gets the ratio $\frac{ p + N_t^C }{ N_t^C N } \approx \frac{1}{N}$, which basically scales with the macrotime scale dimension, as already discussed in \cite{pgd-multiscale}.
	
	The performed numerical tests have shown that the data-driven based solver (figure \ref{fig:solving-scheme}) significantly accelerates the classic one (figure \ref{fig:simulation-up-to-K}). In detail, a speed-up of 2.3 approximately is observed in the evaluation of the nonlinear constitutive relations. Moreover, the overall solver benefits of an additional 0.6 speed-up guaranteed by the separated structure of the predicted right-hand-side (allowing a faster assembly and solution within the MT-PGD solver). Given the algorithm scalability, the computational time and storage gains are further noticeable when more cycles, larger domains and finer meshes are considered, making the procedure attractive in the context of fatigue analyses.
	
	Current research is dealing with variable amplitude loading analyses \cite{variable-amplitude-loadings} and complex loading scenarios as encountered in seismic engineering \cite{pgd-multiscale-2}. Moreover, further developments of this work consider cumulative fatigue damage assessments \cite{cumulative-fatigue-damage,cumulative-fatigue-damage-1}, crack initiation and failure propagation \cite{crack-fatigue}, the usage of the procedure within cyclic-jumping approaches \cite{cycle-jump-0,cycle-jump-1}, cyclic visco-elasto-plastic fatigue problems \cite{viscoelastic,viscoelastic-1} and the extension to many time scales \cite{viscoelastic-2}.
	
    \phantomsection
    \section*{References}
    \addcontentsline{toc}{section}{References}
    \printbibliography[heading=none]
    
        
\end{document}